\ifdefined\venue\else\def\venue{ieee}\fi
\ifdefined\cameraready\else\def\cameraready{no}\fi
\def\venuedefaultieee{ieee}
\ifdefined\showrev\else\ifx\venue\venuedefaultieee\def\showrev{no}\else\def\showrev{yes}\fi\fi

\def\venueieee{ieee}
\ifx\venue\venueieee
  \documentclass[conference]{IEEEtran}
  \IEEEoverridecommandlockouts
  \usepackage[numbers,sort&compress]{natbib}
\else
  \documentclass{article} 
  \usepackage{iclr2027_conference,times}
\fi
\newif\ifieee
\ifx\venue\venueieee\ieeetrue\fi
\newif\ifnamed
\def\venueyes{yes}
\ifieee\namedtrue\else\ifx\cameraready\venueyes\namedtrue\fi\fi

\newcommand{\kernelcount}{689}
\newcommand{\releasecount}{722}

\usepackage{amsmath,amsfonts,bm}

\def\eqref#1{equation~\ref{#1}}

\def\1{\bm{1}}

\DeclareMathAlphabet{\mathsfit}{\encodingdefault}{\sfdefault}{m}{sl}
\SetMathAlphabet{\mathsfit}{bold}{\encodingdefault}{\sfdefault}{bx}{n}

\usepackage{float}
\usepackage{hyperref}
\hypersetup{hidelinks}
\usepackage{url}
\usepackage{xurl}
\usepackage{microtype}
\usepackage{amsmath}
\usepackage{amssymb}
\usepackage{graphicx}
\usepackage{booktabs}
\usepackage{xcolor}
\usepackage{multirow}
\usepackage{comment}
\usepackage{makecell}
\usepackage[inline]{enumitem}
\setlist{topsep=-4pt, partopsep=0pt, itemsep=1pt, parsep=0pt}
\makeatletter
\g@addto@macro\normalsize{%
  \setlength\abovedisplayskip{4pt plus 2pt minus 2pt}%
  \setlength\belowdisplayskip{4pt plus 2pt minus 2pt}%
  \setlength\abovedisplayshortskip{0pt plus 2pt}%
  \setlength\belowdisplayshortskip{2pt plus 2pt minus 1pt}}
\makeatother
\usepackage{bm}
\usepackage{listings}
\usepackage{wrapfig}
\usepackage{capt-of}

\definecolor{codekw}{HTML}{6639BA}    
\definecolor{codeann}{HTML}{0550AE}   
\definecolor{codearr}{HTML}{953800}   
\definecolor{codenum}{HTML}{098658}   
\definecolor{codestr}{HTML}{0A3069}   
\definecolor{codecom}{HTML}{8250DF}   
\definecolor{codetxt}{HTML}{1F2328}   
\definecolor{codeln}{HTML}{8C959F}   

\newcommand{\codedigit}[1]{{\color{codenum}#1}}

\lstdefinestyle{codekernel}{
    language=Python,
    morekeywords={forall,doall,par},
    alsoletter={@_},
    morekeywords=[2]{@scan,@dot,@reduce,scan,dot,reduce},
    morekeywords=[3]{A,B,C,a,a_next,acc,b,c,col,d,e,ip,row_ptr,seam,snap,sum,val,x,y,z},
    literate={0}{{\codedigit{0}}}1 {1}{{\codedigit{1}}}1 {2}{{\codedigit{2}}}1
             {3}{{\codedigit{3}}}1 {4}{{\codedigit{4}}}1 {5}{{\codedigit{5}}}1
             {6}{{\codedigit{6}}}1 {7}{{\codedigit{7}}}1 {8}{{\codedigit{8}}}1
             {9}{{\codedigit{9}}}1,
    numbers=left,
    numberstyle=\tiny\color{codeln},
    stepnumber=1,
    numbersep=5pt,
    xleftmargin=1.5em,
    keepspaces=true,
}

\lstdefinestyle{codec}{
    language=[ANSI]C,
    alsoletter={@_},
    keepspaces=true,
    columns=flexible,
    literate={int64_t}{{{\color{codekw}\bfseries int64\_t}}}7
             {int32_t}{{{\color{codekw}\bfseries int32\_t}}}7
             {size_t}{{{\color{codekw}\bfseries size\_t}}}6
             {__m512d}{{{\color{codekw}\bfseries \_\_m512d}}}7,
    morekeywords={int64_t,size_t,int32_t,uint64_t,bool,restrict,__restrict__,inline},
    morekeywords=[2]{@scan,@dot,@reduce,scan,dot,reduce,pragma,omp,parallel,simd,num_threads,reduction,collapse,schedule},
    morekeywords=[3]{A,B,C,a,a_next,acc,b,c,col,d,e,ip,row_ptr,seam,snap,sum,val,x,y,z,
                     ap,bp,out_index,out_value,va,vb,vc,vd,ve},
    directivestyle=\color{codekw}\bfseries,
    emph={_mm512_mul_pd,_mm512_fmadd_pd,_mm512_stream_pd,_mm512_storeu_pd,_mm512_loadu_pd},
    emphstyle=\color{codeann},
    basicstyle=\ttfamily\footnotesize\color{codetxt},
    numbers=left,
    numberstyle=\tiny\color{codeln},
    stepnumber=1,
    numbersep=4pt,
    xleftmargin=1.4em,
}

\lstnewenvironment{pseudocode}[1][]
{\leavevmode\vspace{0.5em}\lstset{style=codekernel,#1}}
{\vspace{0.5em}}

\ifieee\newcommand{\eqbreak}{\\}\else\newcommand{\eqbreak}{\qquad\qquad}\fi
\newlength{\colwidth}

\newcommand{\fitcolumn}[1]{\ifieee\resizebox{\columnwidth}{!}{#1}\else#1\fi}
\newlength{\widewidth}
\newif\ifwidepack
\ifieee
  \newenvironment{widefigure}[1][]{\ifwidepack\begin{figure*}[!tp]\else\begin{figure*}[t]\fi}{\end{figure*}}
  \newenvironment{widetable}[1][]{\ifwidepack\begin{table*}[!tp]\else\begin{table*}[t]\fi}{\end{table*}}
  \renewenvironment{wrapfigure}[3][]{\begin{figure}[!ht]}{\end{figure}}
  \renewenvironment{wraptable}[3][]{\begin{table}[!ht]}{\end{table}}
\else
  \newenvironment{widefigure}[1][t]{\begin{figure}[#1]}{\end{figure}}
  \newenvironment{widetable}[1][t]{\begin{table}[#1]}{\end{table}}
\fi
\ifieee
  \makeatletter
  \let\ieeemakecaption\@makecaption
  \long\def\@makecaption#1#2{\ifx\@captype\@IEEEtablestring
      \@IEEEfigurecaptionsepspace
      \setbox\@tempboxa\hbox{\normalfont\footnotesize #1.\nobreakspace\nobreakspace #2}%
      \ifdim\wd\@tempboxa>\hsize
        \parbox[t]{\hsize}{\normalfont\footnotesize\noindent #1.\nobreakspace\nobreakspace #2}%
      \else
        \hbox to\hsize{\hfil\box\@tempboxa\hfil}%
      \fi
    \else\ieeemakecaption{#1}{#2}\fi}
  \makeatother
\fi
\graphicspath{{figures/}}

\definecolor{revblue}{HTML}{1F4FD8}
\def\revyes{yes}
\ifx\showrev\revyes
  \newcommand{\rev}[1]{{\color{revblue}#1}}
  \newcommand{\revstart}{\begingroup\color{revblue}}
\else
  \newcommand{\rev}[1]{#1}
  \newcommand{\revstart}{\begingroup}
\fi
\newcommand{\revend}{\endgroup}
\newenvironment{revision}{\revstart}{\revend}

\definecolor{pendred}{HTML}{C62828}
\ifx\showrev\revyes
  
  \newenvironment{pendingblock}{\begingroup\color{pendred}}{\endgroup}
\else

\fi

\title{HPCAgent-Bench: A Framework for Sound Benchmarking of Code-Optimizing Agents, Harnesses, and Skills}

\ifnamed
\ifieee
\author{%
\IEEEauthorblockN{Yakup Koray Budanaz\IEEEauthorrefmark{1},
Lorenzo Paleari\IEEEauthorrefmark{1},
Marcin Copik\IEEEauthorrefmark{1},
Alexandru Calotoiu\IEEEauthorrefmark{1},
Timo Schneider\IEEEauthorrefmark{1},\\
Benjamin Chen Ming Choong\IEEEauthorrefmark{2},
Shriram Chandran\IEEEauthorrefmark{1},
Lars Hulsbergen\IEEEauthorrefmark{1},
Levi Widmer\IEEEauthorrefmark{1},
Torsten Hoefler\IEEEauthorrefmark{1}}
\IEEEauthorblockA{\IEEEauthorrefmark{1}Department of Computer Science, ETH Zurich, Switzerland\\
\{yakupkoray.budanaz, lorenzo.paleari, marcin.copik, alexandru.calotoiu,\\
timo.schneider, shriram.chandran, lars.hulsbergen,\\
levi.widmer, torsten.hoefler\}@inf.ethz.ch}
\IEEEauthorblockA{\IEEEauthorrefmark{2}Institute of High Performance Computing (IHPC), A*STAR, Singapore\\
benjamin.choong@ihpc.a-star.edu.sg}%
}
\else
\author{Yakup Koray Budanaz$^{1}$, Lorenzo Paleari$^{1}$, Marcin Copik$^{1}$, Alexandru Calotoiu$^{1}$,
Timo Schneider$^{1}$, Benjamin Chen Ming Choong$^{2}$, Shriram Chandran$^{1}$, Lars Hulsbergen$^{1}$,
Levi Widmer$^{1}$, Torsten Hoefler$^{1}$ \\
$^{1}$Department of Computer Science, ETH Zurich, Switzerland \\
$^{2}$Institute of High Performance Computing (IHPC), A*STAR, Singapore \\
\texttt{\{yakupkoray.budanaz, lorenzo.paleari, marcin.copik, alexandru.calotoiu, timo.schneider,} \\
\texttt{shriram.chandran, lars.hulsbergen, levi.widmer, torsten.hoefler\}@inf.ethz.ch} \\
\texttt{benjamin.choong@ihpc.a-star.edu.sg}}
\fi

  \ifieee\else\iclrfinalcopy\fi
\else
  \author{Anonymous authors\\Paper under double-blind review}
\fi
\newcommand{\figcodefont}{\ttfamily\ifieee\footnotesize\else\scriptsize\fi}

\newcommand{\artifacturl}{\url{https://doi.org/10.5281/zenodo.22974128}}
\newcommand{\githuburl}{\url{https://github.com/spcl/HPCAgent-Bench}}

\newcommand{\codefoot}[1]{\par\vspace{2pt}{\centering\small #1\par}}


\makeatletter
\DeclareRobustCommand{\_}{\ifmmode\nfss@text{\textunderscore}\else\textunderscore\fi\allowbreak}
\makeatother
\begin{document}

\maketitle
\ifieee\thispagestyle{plain}\pagestyle{plain}\fi

\begin{abstract}
Coding agents are increasingly used to optimize code, but existing benchmarks fix one accelerator, programming model, or harness, so they cannot isolate what each component contributes or show whether added skills and tools repay their cost. We present HPCAgent-Bench with \rev{\releasecount{}} kernels in three tracks: loop-level optimization challenges, kernels extracted from HPC codes, and machine-learning models. The framework specifies each kernel once in NumPy, derives C and Fortran references, accepts submissions through a common C ABI or Python interface, and runs them on CPUs or GPUs, on one or multiple nodes. An external judge checks correctness on hidden fuzzed inputs and times each submission, closing the paths agents use to exploit the testing system. We introduce \rev{a reporting protocol,} \textit{intervention efficacy}, \rev{which reports} an intervention, such as adding a tool, through paired runs on speedup, success rate, and cost, and \rev{a metric,} \textit{weighted token cost}, \rev{which} reports input, cached-input, and output tokens under explicit weights. Agents reach up to \rev{$5.0\times$} speedup over automatic parallelizers, and their optimizations transfer to another machine: $93$--$99\%$ of answers without device-specific code remain correct, and per-kernel speedups keep a rank correlation of $r_s\ge0.84$. \rev{Of 48 intervention comparisons, 6 change speed and 9 change token cost significantly, and 5 leave speed equivalent.} Because token weighting decides whether a saving is significant, we state the weights and release every token count as a standard for cost reporting. \rev{HPCAgent-Bench is a broad, hardened benchmark for the sound evaluation of agentic optimization, with transparent cost metrics.}
\end{abstract}

\ifieee
\begin{IEEEkeywords}
coding agents, code optimization, benchmarking, high-performance computing, large language models
\end{IEEEkeywords}
\fi

\section{Introduction}
\label{sec:intro}

Optimizing code for performance is labor-intensive. Each target architecture demands its own cycle of profiling and tuning. Auto-tuning and performance-portability frameworks such as Halide~\citep{ragankelley2013halide}, TVM~\citep{chen2018tvm}, and DaCe~\citep{bennun2019sdfg}, and programming models such as OpenMP~\citep{openmp50} and Kokkos~\citep{trott2022kokkos} shorten that cycle. LLMs rewrite programs for speed from examples of faster code~\citep{shypula2024pie}, and agentic systems now run the optimization loop end to end~\citep{novikov2025alphaevolve}.

The benchmarks used to evaluate these agents are narrow and their measurements are contested. Existing benchmarks target one device~\citep{li2025tritonbench,zhang2026ptxbench,chen2025pcebench}, support a restricted set of languages or programming models~\citep{ouyang2025kernelbench,nichols2024pareval}, or cover only one domain, such as ML models, BLAS routines, or Python libraries~\citep{ouyang2025kernelbench,zang2026kernelgenbench,he2025sweperf,press2025algotune}. None spans low-level and parallelism challenges, production HPC code, and ML models (Table~\ref{tab:capabilities-matrix}). Reported speedups are often unreliable: a re-evaluation of four benchmarks found that only $6.11$\% of implementations labeled as performant were significantly faster than their canonical implementations when re-timed over 30 runs~\citep{le2026rethinking}, replayed reference patches lose their speedup on other machines~\citep{chen2026reproducibilityaudit}, and agents exploit evaluation loopholes~\citep{lange2025robustkbench} and overfit to exposed tests and static inputs~\citep{zhang2026kernelbenchverifiedllmgeneratedkernelsactually,shetty2025gso} (\S\ref{sec:related}).

\providecommand{\cmark}{\textcolor{green!60!black}{\scalebox{1.25}{\ensuremath{\bm{\checkmark}}}}}
\providecommand{\xmark}{\textcolor{red!85!black}{\scalebox{1.25}{\ensuremath{\bm{\times}}}}}

\begin{widetable}[!t]
\centering
\renewcommand{\arraystretch}{1}
\setlength{\tabcolsep}{3pt}
\small
\resizebox{\textwidth}{!}{%
\begin{tabular}{@{}lcccccccccccc@{}}
\toprule
 & & & & & & & \multicolumn{5}{c}{\rev{\textbf{\scriptsize Protocol}}} \\
\cmidrule(l){8-12}
\textbf{\small Benchmark} &
\makecell{\textbf{\scriptsize Low-level /}\\\textbf{\scriptsize Parallelism}\\\textbf{\scriptsize Reasoning}} &
\makecell{\textbf{\scriptsize HPC}\\\textbf{\scriptsize Proxy-Apps}} &
\rev{\makecell{\textbf{\scriptsize ML}\\\textbf{\scriptsize Kernels}}} &
\makecell{\textbf{\scriptsize Multi-}\\\textbf{\scriptsize Node}} &
\rev{\textbf{\scriptsize Devices}} &
\textbf{\scriptsize Baselines} &
\rev{\makecell{\textbf{\scriptsize Hidden}\\\textbf{\scriptsize inputs}}} &
\rev{\makecell{\textbf{\scriptsize Held-out}\\\textbf{\scriptsize final grade}}} &
\rev{\makecell{\textbf{\scriptsize Tested}\\\textbf{\scriptsize comparisons}}} &
\rev{\makecell{\textbf{\scriptsize Interval}}} &
\rev{\makecell{\textbf{\scriptsize Cost}}} \\
\midrule

\textbf{HPCAgent-Bench (ours)}
    & \cmark & \cmark & \rev{\cmark} & \cmark & \rev{C+G} & U, F & \rev{\cmark} & \rev{\cmark} & \rev{\cmark} & \rev{\cmark} & \rev{\cmark} \\

ParEval~\citep{nichols2024pareval}
    & \rev{\cmark} & \xmark & \rev{\xmark} & \cmark & \rev{C+G} & U & \rev{\xmark} & \rev{\xmark} & \rev{\xmark} & \rev{\xmark} & \rev{\cmark} \\

AlgoTune~\citep{press2025algotune}
    & \xmark & \xmark & \rev{\xmark} & \xmark & \rev{C} & F & \rev{\cmark} & \rev{\cmark} & \rev{\xmark} & \rev{\xmark} & \rev{\cmark} \\

KernelBench~\citep{ouyang2025kernelbench}
    & \xmark & \xmark & \rev{\cmark} & \xmark & \rev{G} & F & \rev{\xmark} & \rev{\xmark} & \rev{\xmark} & \rev{\xmark} & \rev{\xmark} \\

TritonBench~\citep{li2025tritonbench}
    & \xmark & \xmark & \rev{\cmark} & \xmark & \rev{G} & E, F & \rev{\xmark} & \rev{\xmark} & \rev{\xmark} & \rev{\xmark} & \rev{\xmark} \\

MultiKernelBench~\citep{wen2025multikernelbench}
    & \xmark & \xmark & \rev{\cmark} & \xmark & \rev{G+O} & F & \rev{\xmark} & \rev{\xmark} & \rev{\xmark} & \rev{\xmark} & \rev{\xmark} \\

robust-kbench~\citep{lange2025robustkbench}
    & \xmark & \xmark & \rev{\cmark} & \xmark & \rev{G} & F & \rev{\xmark} & \rev{\xmark} & \rev{\xmark} & \rev{\xmark} & \rev{\cmark} \\

PCEBench~\citep{chen2025pcebench}
    & \xmark & \xmark & \rev{\xmark} & \xmark & \rev{C} & U & \rev{\xmark} & \rev{\xmark} & \rev{\xmark} & \rev{\xmark} & \rev{\xmark} \\

SWE-Perf~\citep{he2025sweperf}
    & \xmark & \xmark & \rev{\xmark} & \xmark & \rev{C} & U, E & \rev{\xmark} & \rev{\xmark} & \rev{\xmark} & \rev{\xmark} & \rev{\xmark} \\

SWE-fficiency~\citep{ma2026sweefficiency}
    & \xmark & \xmark & \rev{\xmark} & \xmark & \rev{C} & U, E & \rev{\xmark} & \rev{\xmark} & \rev{\xmark} & \rev{\cmark} & \rev{\cmark} \\

GSO~\citep{shetty2025gso}
    & \cmark & \xmark & \rev{\xmark} & \xmark & \rev{C} & U, E & \rev{\cmark} & \rev{\cmark} & \rev{\xmark} & \rev{\cmark} & \rev{\xmark} \\

KernelGenBench~\citep{zang2026kernelgenbench}
    & \xmark & \xmark & \rev{\cmark} & \xmark & \rev{G+O} & E, F & \rev{\xmark} & \rev{\xmark} & \rev{\xmark} & \rev{\cmark} & \rev{\cmark} \\

CUDAHercules~\citep{li2026cudahercules}
    & \xmark & \cmark & \rev{\cmark} & \xmark & \rev{G} & E, F & \rev{\xmark} & \rev{\xmark} & \rev{\xmark} & \rev{\xmark} & \rev{\xmark} \\

PerfCodeBench~\citep{jing2026perfcodebench}
    & \xmark & \xmark & \rev{\cmark} & \xmark & \rev{C+G} & U, E & \rev{\cmark} & \rev{\xmark} & \rev{\xmark} & \rev{\xmark} & \rev{\cmark} \\

FormulaCode~\citep{sehgal2026formulacode}
    & \xmark & \xmark & \rev{\xmark} & \xmark & \rev{C} & U, E & \rev{\xmark} & \rev{\xmark} & \rev{\xmark} & \rev{\xmark} & \rev{\cmark} \\

CommBench~\citep{ma2026commbench}
    & \cmark & \xmark & \rev{\cmark} & \cmark & \rev{G} & E & \rev{\xmark} & \rev{\xmark} & \rev{\xmark} & \rev{\xmark} & \rev{\cmark} \\

PTXBench~\citep{zhang2026ptxbench}
    & \cmark & \xmark & \rev{\cmark} & \xmark & \rev{G} & F & \rev{\xmark} & \rev{\xmark} & \rev{\xmark} & \rev{\cmark} & \rev{\cmark} \\

\rev{ParallelKernelBench~\citep{chan2026parallelkernelbench}}
    & \rev{\cmark} & \rev{\xmark} & \rev{\cmark} & \rev{\xmark} & \rev{G} & \rev{F} & \rev{\xmark} & \rev{\xmark} & \rev{\xmark} & \rev{\xmark} & \rev{\xmark} \\

\rev{D2K-Bench~\citep{li2026d2kbench}}
    & \rev{\cmark} & \rev{\xmark} & \rev{\cmark} & \rev{\xmark} & \rev{G} & \rev{E, F} & \rev{\cmark} & \rev{\xmark} & \rev{\xmark} & \rev{\xmark} & \rev{\cmark} \\

\rev{KB-Verified~\citep{zhang2026kernelbenchverifiedllmgeneratedkernelsactually}}
    & \rev{\xmark} & \rev{\xmark} & \rev{\cmark} & \rev{\xmark} & \rev{G} & \rev{F} & \rev{\cmark} & \rev{\xmark} & \rev{\xmark} & \rev{\xmark} & \rev{\xmark} \\

\bottomrule
\end{tabular}}
\vspace{-0.5em}
\caption{\label{tab:capabilities-matrix}
Capabilities of agentic code-optimization benchmarks (definitions in Appendix~\ref{app:capabilities}). \rev{Ours: one problem accepts any language behind a shared C ABI. KB-Verified: KernelBench-Verified. \emph{Devices}: \textbf{C} CPU, \textbf{G} GPU, \textbf{O} other accelerators.} \emph{Baselines}: \textbf{U} sequential, \textbf{F} library-optimized, \textbf{E} expert-written.}
\end{widetable}

\rev{HPCAgent-Bench has one objective: a sound comparison of agent setups that shows which results transfer across tracks, languages, devices, and models, and which do not. Such a comparison requires one benchmark under one shared protocol for every track.} An experiment fixes the kernel and the agent's model, language, device, harness, skills, and tools; an intervention changes exactly one of them, and we compare speedup and token cost with the unchanged setup (\S\ref{sec:score}). The \rev{\releasecount{}} kernels form three tracks (Table~\ref{tab:corpus}): \textbf{loop-level reasoning}, isolating classical compiler transformations; \textbf{scientific computing}, with microkernels and proxy applications from production codes and benchmark suites, organized by the thirteen Berkeley dwarfs~\citep{asanovic2006berkeleyview}; and \textbf{machine learning}, with deep-learning operators and models. \rev{Three requirements shaped the design. (a)~Existing suites must become agent tasks: because compiler test suites such as TSVC-2~\citep{maleki2011tsvc2} and PolyBench~\citep{pouchet2012polybench} run on fixed arrays and check only a checksum, each kernel gets a NumPy reference, a seeded input generator, and a C ABI and Python interface. (b)~Inputs and tolerance must be designed together: fresh inputs prevent cached answers, and the tolerance must accept sound floating-point reorderings but reject wrong outputs, a separation that cancellation made impossible on some original inputs. (c)~One judge must grade all three tracks, on one node or across ranks, in whatever precision and data layout a submission chooses, because \rev{results compare} across tracks only when the same judge grades them.} \rev{Each of our contributions serves that objective:}

\begin{enumerate}[label=(\arabic*), itemjoin={{; }}, itemjoin*={{; and }}, itemsep=1pt, topsep=1pt, parsep=0pt]

\item A \textbf{benchmarking framework} for coding agents on kernel optimization with controlled interventions: an external judge grades correctness on hidden fuzzed inputs and times every submission, so agents cannot hard-code answers or report their own times; a speedup counts only when it is statistically significant; a result cache hidden from agents makes grading fast; and runs use fixed container images with isolated execution (\S\ref{sec:design}, Appendix~\ref{app:protocol}).

\item \rev{A new \textbf{reporting protocol}, \textbf{\emph{intervention efficacy}}, that fixes which comparisons run, which statistical tests decide each comparison's verdict, and what the report contains (\S\ref{sec:score}),} and a \textbf{weighted token cost} over input, cached-input, and output tokens that covers the cost conventions of prior work (\S\ref{sec:score}).

\item A \textbf{corpus} of \rev{\releasecount{}} kernels that combines HPC proxy applications, loop-level reasoning challenges, and ML models, unified under one ABI for both CPU and GPU execution (\rev{Tables~\ref{tab:corpus} and~\ref{tab:selection}})\rev{: 494 kernels adapted from existing benchmarks into agent tasks (201 of them with repaired inputs) and 228 new ones (181 written for the benchmark, 47 extracted from production codes and proxy applications)}.

\item An \textbf{evaluation} of the framework across eight interventions, three open-weight models, two devices, and a second machine (\S\ref{sec:evaluation}).
\end{enumerate}

\section{Design}
\label{sec:design}

The rest of this section follows Figure~\ref{fig:overview}. From each kernel's NumPy specification the framework generates the C and Fortran references, the language bindings, and the prompt. An episode has one prompt, and its answer is one artifact: a shared library exporting the kernel's C-ABI symbol, a source file, or a Python binding. Containers run under any runtime compliant with the Open Container Initiative (OCI), Docker by default (Appendix~\ref{app:protocol}).

\subsection{What a run declares}
\label{sec:config}

\begin{widefigure}[!t]
\centering
\vspace{-0em}
\includegraphics[width=\widewidth]{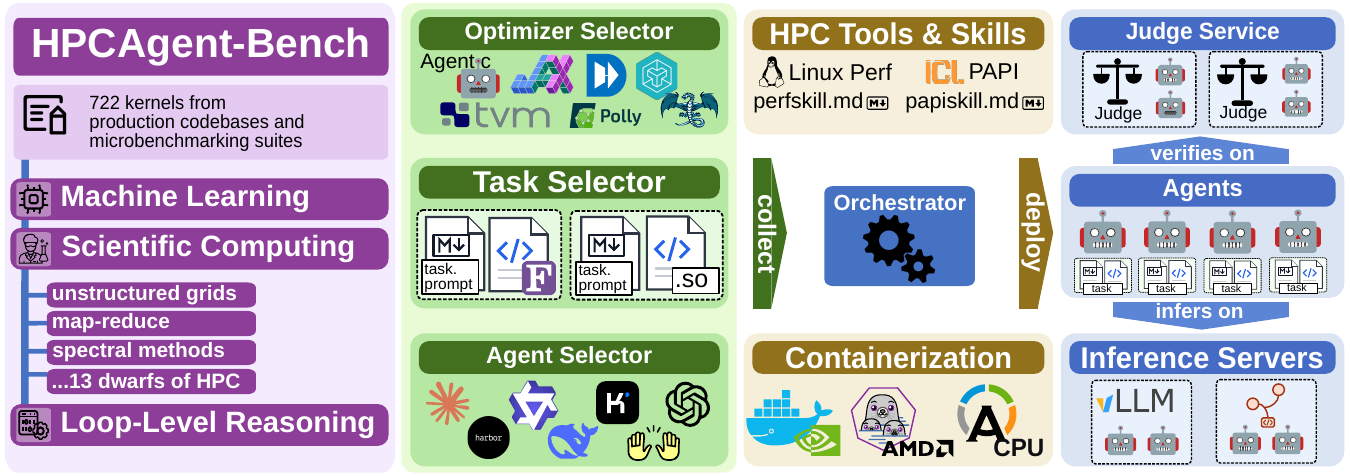}
\vspace{-1em}
\caption{Overview of HPCAgent-Bench. Selectors, fields of the run's configuration, fix the optimizer, the tasks, and the agent with its skills and tools. The orchestrator, the part of the framework that schedules episodes, runs one agent per task. \rev{The judge checks each candidate's speedup and validity.}}
\label{fig:overview}
\vspace{-0.0em}
\end{widefigure}

\textbf{Per kernel}: Each kernel comes with a manifest, a YAML file that lists its size parameters, its input and output arrays, and its difficulty level, plus optional tuning knobs. \textbf{Per run}: A run's \texttt{config.yaml} selects the kernels and fixes the language mode (e.g., \texttt{C} or \texttt{any}), the submission mode (\S\ref{sec:submission-modes}), the floating-point precision, the array residency (\texttt{host} or \texttt{device}) and the optimizer: an LLM agent with its harness, skills, and tools\rev{, or an automatic parallelizer such as Pluto or PPCG}. Residency sets the timing boundary. A \texttt{device} run receives device pointers, so host--device transfers stay outside the timed sample, and the judge synchronizes before it records a runtime (Appendix~\ref{app:protocol}).

\textbf{Scaling.} The framework supports strong- and weak-scaling experiments over the number of processing elements (PEs) $P$, on one node or several. Weak scaling grows the manifest's declared sizes with $P$ so each PE keeps the base work, while strong scaling keeps total work constant (Appendix~\ref{app:distributed}).

\subsection{Inputs, secrecy, and verification}
\label{sec:inputs}

\textbf{Inputs.}\label{sec:datavalidity}\label{sec:structural} Each input draw takes the kernel's sizes from its declared range under the manifest's constraints and cycles the array contents through uniform, normal, and lognormal distributions, sizes and contents both drawn from \rev{one of two seeds}: the secret score seed for the score tool and the secret submission seed for the final grade. A kernel defined only on constrained inputs (such as one requiring positive-definite matrices) declares a value domain or a custom \texttt{initialize} function.

\textbf{Fuzzing and secrecy.}\label{sec:fuzz} Correctness is evaluated on fuzzed inputs generated from two independent secret seeds. The \texttt{/score} tool evaluates a candidate using inputs generated from the secret score seed, while \texttt{/submit} evaluates the answer using inputs from the secret submission seed. This prevents the agent from optimizing specifically for the inputs it has already observed. \rev{The two seeds generate disjoint inputs, and each held-out evaluation generates new values.}

\textbf{Submission modes.}\label{sec:submission-modes} A run fixes how often an agent may call each route: \emph{open-submission} (unlimited scores and submissions), \emph{single-submission} (unlimited scores, one submission) or \emph{blind} (no scores, one submission). We recommend single-submission for comparing optimizers.

\textbf{Numerical agreement.}\label{sec:verify} Correctness is all-or-nothing. Integer and boolean outputs must match exactly; a floating-point output must satisfy $|x - x_{\text{ref}}| \le \texttt{atol} + \texttt{rtol}\,|x_{\text{ref}}|$ elementwise, with \texttt{rtol} and \texttt{atol} fixed per precision. An optimization may reorder a sum and change its rounding, so \texttt{atol} is raised to at least $\varepsilon_{\text{acc}}\sqrt{\ell}\,\lVert x_{\text{ref}}\rVert_\infty$: for a sum of $\ell$ terms into one output element ($\ell=1$ for elementwise kernels), independent signed rounding errors grow as $\sqrt{\ell}$ with high probability, not $\ell$~\citep{higham2019probabilistic}, where $\varepsilon_{\text{acc}}$ is the machine epsilon of the precision the sum accumulates in~\citep{blanchard2020mixed} (Appendix~\ref{app:protocol}). \rev{Errors small enough to pass each element's band can still add up across the output, so an $L_2$ check also bounds the error of the whole output, $\lVert x-x_{\text{ref}}\rVert_2\le\texttt{rtol}\,\lVert x_{\text{ref}}\rVert_2$ (Table~\ref{tab:checks}).}

\rev{\textbf{Extensibility.} \emph{Adding a kernel} takes a NumPy reference and a manifest, \emph{adding a machine} a system description and a container image, and \emph{adding an optimizer} one plugin, with no change to the benchmark or judge. Each of our three plugins took 4--8 person-hours: one exposes DaCe's code transformations as agent tools~\citep{bennun2019sdfg}, one runs Optimas~\citep{wu2025optimas} to tune the agent's instructions across rounds, and one wraps the judge in a Gymnasium environment~\citep{towers2024gymnasium} for optimizers that are not agents; all three run under the same limits and the single-submission rule (Appendix~\ref{app:selection}).}

\subsection{Corpus sources}
\label{sec:selection}

The corpus holds the three tracks (Table~\ref{tab:corpus}). Loop-Level Reasoning (\textbf{LLR}) challenges descend from the Test Suite for Vectorizing Compilers, TSVC-2~\citep{callahan1988tsvc,maleki2011tsvc2}, modernized and extended with adversarial variants. Scientific Computing (\textbf{SC}) \rev{has 179 kernels: 47} extracted from production codes in weather and climate~\citep{zaengl2015icon,putman2007fv3,afanasyev2021gridtools}, electronic structure~\citep{giannozzi2009qe,borstnik2014dbcsr,wang2008ls3df,luisier2006omen}, plasma physics and fusion energy~\citep{myers2021warpx,dudson2009bout}, computational seismology~\citep{uphoff2020seissol,petersson2015sw4} and molecular dynamics~\citep{thompson2022lammps,pall2013gromacs} and from HPC proxy applications~\citep{tramm2014xsbench,karlin2013lulesh}\rev{, 75 ported from public suites and codes, and 57 written for the benchmark}; we organize them by the thirteen Berkeley dwarfs~\citep{asanovic2006berkeleyview}. \rev{Of the ported kernels, 32 come from PolyBench/C~\citep{pouchet2012polybench}, 15 from NPBench~\citep{ziogas2021npbench}, 8 from Rodinia~\citep{che2009rodinia}, 6 from OpenDwarfs~\citep{krommydas2016opendwarfs}, and 14 from other public codes.} Machine Learning (\textbf{ML}) models range from deep-learning primitives to full networks, mainly adapted from KernelBench~\citep{ouyang2025kernelbench}. No existing benchmark provides comparable breadth across kernel types and application domains (Table~\ref{tab:capabilities-matrix}, Table~\ref{tab:selection}).
\section{Scoring}
\label{sec:score}
\ifieee\begin{widetable}[ht]
\centering\scriptsize
\setlength{\tabcolsep}{3pt}
\begin{tabular}{@{}llp{0.32\textwidth}lrr@{}}
\toprule
Experiment & Kernels & Languages (device) & Models & Tokens & \rev{Output} \\
\midrule
LLR-Focus40 & 40 LLR (Focus40) & C, Fortran (CPU); HIP, Triton \rev{(GPU)} & Q, K, O & \rev{50M} & \rev{12M} \\
\rev{CPF} & \rev{the same 40} & \leavevmode\rev{C (CPU)} & \rev{Q, K, O} & \rev{12M} & \rev{none} \\
\rev{Blind vs.\ Open} & the same 40 & \leavevmode\rev{C (CPU)} & \rev{K, O} & \rev{50M} & \rev{12M} \\
Git-SciComp & \rev{27 SC (in SciComp40)} & C (CPU) & Q, K, O & \rev{50M} & \rev{12M} \\
\rev{SciComp40} & \rev{40} SC & C (CPU)\rev{; Fortran (CPU), HIP (GPU): Q, O} & Q, K, O & \rev{50M} & \rev{12M} \\
Harness20 & 14 SC, 6 LLR & C (CPU) & Q, O & \rev{48M} & \rev{12M} \\
\rev{Blind vs.\ Single} & \rev{Focus40; SciComp40} & \leavevmode\rev{C (CPU); on Focus40 also Fortran (CPU; K) and HIP (GPU; Q, O)} & \rev{Q, K, O} & \rev{50M} & \rev{12M} \\
MLScale & \rev{20} distributed ML models & HIP with RCCL (1--16 GPUs, 4 nodes) & Q, K, O & \rev{50M} & \rev{12M} \\
\bottomrule
\end{tabular}
\caption{Experiments. Focus40: the 40-kernel LLR focus set; Q, K, O: Qwen3.8-27B, Kimi-K2.7, OSS-120B. Per-episode limits\rev{, which every counted answer stayed within}: \emph{Tokens}, total tokens (\S\ref{sec:token-cost}; every turn's input counts in full, cached part included); \rev{\emph{Output}, output tokens. Episodes also had a wall-time limit; it is not compared across setups, as serving speed differs by model (Kimi-K2.7 needed up to 32\,h for the same tokens).} \rev{Exceptions in Table~\ref{tab:exp-setups} and Appendix~\ref{app:budgets}.}}
\label{tab:experiments}
\end{widetable}
\fi

\textbf{Speedup score.}\label{sec:metric} An answer to task $i$ is graded for correctness on a broad set of fuzzed inputs and timed against its baseline on $m$ inputs that differ in size and, where the kernel has control-flow flags, in flag setting. A task is \emph{solved} if every graded input is correct. On each input, the baseline and the answer run once as a warmup and then $n$ times, so an answer cannot hardcode or overfit to one input (\S\ref{sec:evasion}). On input $j\in[m]$, the speedup $s_{ij}$ is the baseline's median runtime divided by the answer's. It is credited if a one-sided Mann--Whitney $U$ test in the direction of the medians gives $p<\alpha$, with $\alpha=0.1$, so a confirmed slowdown is credited below $1$; it is $1$ otherwise. The task score is $S_i$, the geometric mean of its speedups $s_i=(s_{i1},\dots,s_{im})$,
\begin{equation}
S_i=\operatorname{GM}(s_i),\qquad \operatorname{GM}(x)=\Big(\prod_{k=1}^{|x|} x_k\Big)^{1/|x|}.
\label{eq:speedup-score}
\end{equation}
$S_i$ has no ceiling. An unsolved task has no score. A benchmark run over $N$ tasks with solved set $\mathcal{P}$ reports the \emph{success rate} $R=|\mathcal{P}|/N$ and the \emph{speedup score} $\operatorname{GM}_{i\in\mathcal{P}} S_i$. We compare runtimes with a rank test because they are rarely normal, and a rank test assumes no distribution~\citep{hoefler2015scientific}. Appendix~\ref{app:protocol} gives the timing protocol.

\textbf{Scaling score.}\label{sec:scaling-score} A scaling experiment runs an answer on $P$ PEs and scores its parallel efficiency $\eta_i(P)$ against $B_i$, the runtime of the task's baseline on one PE at the base problem $N_1$ (e.g., PyTorch on one GPU for the ML models), where $T_i(P)$ is the answer's runtime on $P$ PEs:
\begin{equation}
\begin{gathered}
\eta_i(P)=\frac{B_i}{P\,T_i(P)}\quad\text{(strong scaling)},\eqbreak
\eta_i(P)=\frac{r_i(P)\,B_i}{P\,T_i(P)}\quad\text{(weak scaling)},
\end{gathered}
\label{eq:scaling}
\end{equation}
where $r_i(P)$ is the grown problem's work in units of the base problem's ($P$ for exact growth). $\eta_i(P)$ is thus the speedup over the baseline divided by $P$: $1$ matches ideal scaling of the baseline, and above $1$ the answer beats the baseline per PE. We anchor at the baseline, not at the answer's own single-PE run $T_i(1)$, since slow code scales well. A PE count enters only if both runs are correct.

\textbf{Intervention efficacy \rev{protocol}.} An \emph{intervention} adds something to an agent's setup or removes it, such as a skill, a tool, or a new prompt framing. We run the agent without it (before) and with it (after) on the same kernels $\mathcal{K}$, and let $\mathcal{B}\subseteq\mathcal{K}$ be the kernels both runs solve. Three ratios measure the effect: the \emph{success ratio} $\rho_R$ compares the success rates $R$, the \emph{speedup ratio} $\rho_S$ compares the scores $S_i$ over $\mathcal{B}$, and the \emph{cost ratio} $\rho_C$ compares the token costs $C_i$ over $\mathcal{K}$. Each is $1$ for no effect and above $1$ for an improvement:
\begin{equation}
\begin{gathered}
\rho_R=\frac{R^{\text{after}}}{R^{\text{before}}},\eqbreak
\rho_S=\frac{\operatorname{GM}_{i\in\mathcal{B}}\, S_i^{\text{after}}}{\operatorname{GM}_{i\in\mathcal{B}}\, S_i^{\text{before}}},\eqbreak
\rho_C=\frac{\operatorname{GM}_{i\in\mathcal{K}}\, C_i^{\text{before}}}{\operatorname{GM}_{i\in\mathcal{K}}\, C_i^{\text{after}}}.
\end{gathered}
\label{eq:efficacy}
\end{equation}
$\rho_S$ uses only $\mathcal{B}$, since an unsolved kernel has no speedup. $\rho_C$ keeps every kernel, since a failed task still spends tokens. Alongside $\rho_R$ we report $g$ and $l$, the numbers of kernels solved only after and only before the intervention. \rev{Efficacy is a reporting protocol in three parts (full protocol and defaults: Table~\ref{tab:protocol-defaults}, Appendix~\ref{app:efficacy-statistics}). \emph{Design}: the two setups differ only in the intervention, run the same kernels with one scored episode each under the same limits, and are graded on held-out inputs the agent never sees. \emph{Inference}: a paired difference test, an equivalence test, and a multiple-comparison correction decide each verdict; by default, a paired sign-flip permutation test compares speed and cost, the exact McNemar test the solve rate, two one-sided tests~\citep[TOST;][]{schuirmann1987tost} test equivalence within $1.2\times$ for ratios and 10 points for solve rates, and the Benjamini--Hochberg procedure~\citep{benjamini1995fdr} controls the false discovery rate at 5\% within each experiment and measure. \emph{Output}: each ratio comes with its 95\% confidence interval~\citep{agarwal2021precipice,miller2024errorbars} and is read as \emph{different}, \emph{equivalent} (a change as large as the margin is ruled out), or \emph{inconclusive} (neither). We recommend adopting an intervention that is equivalent in solve rate and speed but cheaper, and weighing a speed gain against its cost ratio.}

\textbf{Token cost.}\label{sec:token-cost}
Prior work accounts for token cost in incompatible ways (\S\ref{sec:related}); we represent these conventions with a single weighted sum. From the agent's transcript we read three components: input $T_{\mathrm{in}}$, the prompt tokens not in the previous request; cached input $T_{\mathrm{cache}}$, the prompt tokens that were, all assumed served from the cache; and output $T_{\mathrm{out}}$, thinking tokens included. A weight vector $w$ gives the cost
\begin{equation}
C^{w} = w_{\mathrm{in}} T_{\mathrm{in}} + w_{\mathrm{cache}} T_{\mathrm{cache}} + w_{\mathrm{out}} T_{\mathrm{out}}.
\label{eq:token-cost}
\end{equation}
We use three weightings: \emph{billed}, $w=(1,0.1,1)$ (our default), which charges cached input at a tenth of the input rate, as major providers do~\citep{anthropic2026promptcaching,openai2026promptcaching}; \emph{effective}, $w=(1,0,1)$, which counts every context token once; and \emph{total}, $w=(1,1,1)$, which charges every prompt in full.

\section{Evaluation}
\label{sec:evaluation}

A benchmark for agentic optimization should be valid, portable, controlled, general, and transparent. We evaluate HPCAgent-Bench against these five properties from the view of a user who wants to measure what an agent can optimize. \emph{Validity}: can a reported speedup be trusted (\S\ref{sec:evasion})? \emph{Portability}: does it hold on another machine (\S\ref{sec:transfer})? \emph{Control}: does a change to the agent's setup change what it achieves (\S\ref{sec:interventions})? \emph{Generality}: does the benchmark extend past single-device kernels, to accelerators and distributed computing (\S\ref{sec:targets})? \emph{Transparency}: can a result be traced to the code behind it (\S\ref{sec:llr-behaviour})?

\subsection{Experimental setup}
\label{sec:setup}

\textbf{Setups.} Single-node grades run on one AMD Instinct MI300A APU; MLScale uses $16$ GPUs on four nodes. \rev{Blind vs.\ Open and Harness20 run open-submission; the other experiments, single-submission.} A \emph{setup} fixes model (Qwen3.8-27B, Kimi-K2.7, OSS-120B), language, device, skill packets, and harness (Claude Code by default). Cost uses the billed weights $w=(1,0.1,1)$. \rev{Each setup runs one episode per kernel. GPU speedups, like CPU speedups, are over the CPU baseline; hand-tuned Triton solutions give a GPU reference} (Appendix~\ref{app:setup-details}).

\rev{\textbf{Kernel subsets.} Each subset takes a fixed quota per level of the corpus, drawn at random within each level: 24 level-1 and 16 level-2 LLR kernels (LLR-Focus40), and 3, 10, and 22 SC kernels at levels 1--3 (SciComp35). Most SC comparisons use SciComp40, which drops one SciComp35 kernel and adds six (Tables~\ref{tab:experiments} and~\ref{tab:exp-sc}).}

\ifieee\else\begin{widetable}[ht]
\centering\scriptsize
\setlength{\tabcolsep}{3pt}
\begin{tabular}{@{}llp{0.32\textwidth}lrr@{}}
\toprule
Experiment & Kernels & Languages (device) & Models & Tokens & \rev{Output} \\
\midrule
LLR-Focus40 & 40 LLR (Focus40) & C, Fortran (CPU); HIP, Triton \rev{(GPU)} & Q, K, O & \rev{50M} & \rev{12M} \\
\rev{CPF} & \rev{the same 40} & \leavevmode\rev{C (CPU)} & \rev{Q, K, O} & \rev{12M} & \rev{none} \\
\rev{Blind vs.\ Open} & the same 40 & \leavevmode\rev{C (CPU)} & \rev{K, O} & \rev{50M} & \rev{12M} \\
Git-SciComp & \rev{27 SC (in SciComp40)} & C (CPU) & Q, K, O & \rev{50M} & \rev{12M} \\
\rev{SciComp40} & \rev{40} SC & C (CPU)\rev{; Fortran (CPU), HIP (GPU): Q, O} & Q, K, O & \rev{50M} & \rev{12M} \\
Harness20 & 14 SC, 6 LLR & C (CPU) & Q, O & \rev{48M} & \rev{12M} \\
\rev{Blind vs.\ Single} & \rev{Focus40; SciComp40} & \leavevmode\rev{C (CPU); on Focus40 also Fortran (CPU; K) and HIP (GPU; Q, O)} & \rev{Q, K, O} & \rev{50M} & \rev{12M} \\
MLScale & \rev{20} distributed ML models & HIP with RCCL (1--16 GPUs, 4 nodes) & Q, K, O & \rev{50M} & \rev{12M} \\
\bottomrule
\end{tabular}
\caption{Experiments. Focus40: the 40-kernel LLR focus set; Q, K, O: Qwen3.8-27B, Kimi-K2.7, OSS-120B. Per-episode limits\rev{, which every counted answer stayed within}: \emph{Tokens}, total tokens (\S\ref{sec:token-cost}; every turn's input counts in full, cached part included); \rev{\emph{Output}, output tokens. Episodes also had a wall-time limit; it is not compared across setups, as serving speed differs by model (Kimi-K2.7 needed up to 32\,h for the same tokens).} \rev{Exceptions in Table~\ref{tab:exp-setups} and Appendix~\ref{app:budgets}.}}
\label{tab:experiments}
\end{widetable}
\fi

\begin{revision}
\subsubsection{Run-to-run variation}
\label{sec:aa}
\ifieee\else\noindent\begin{minipage}[c]{0.53\linewidth}\fi
An A/A study measures how much one agent varies between identical runs: \rev{four kernels, among them the ML kernel \texttt{gelu} (Figure~\ref{fig:temperature}),} run 20 times for each of two models at temperature 1 and 0 (Appendix~\ref{app:repeat}). On \texttt{kmp}, Qwen3.8-27B's episodes use four algorithms and reach $0.51$--$100\times$ at temperature 1 \rev{and $1.8$--$11\times$ at temperature 0}. At the default temperature, the standard deviation of log speedup ranges from near $0$ to $1.7$ ($0.45$--$1.03$ in four of \rev{eight} groups), and token counts vary less (standard deviation of log tokens \rev{$0.38$}--$0.71$). \rev{On \texttt{gelu} the spread is small ($0.08$--$0.16$); we attribute this to the PyTorch baseline being parallel already, which leaves agents tuning rather than restructuring.} On 40 kernels at 80\% power, one episode per kernel therefore detects only speed effects of about $1.3$--$2.0\times$ or more and cost effects of about $1.4\times$ or more (Appendix~\ref{app:repeat}).
\ifieee\par\begin{figure}[!tb]\centering\else\end{minipage}\hfill
\begin{minipage}[c]{0.45\linewidth}\centering\fi
\includegraphics[width=\ifieee\colwidth\else\linewidth\fi]{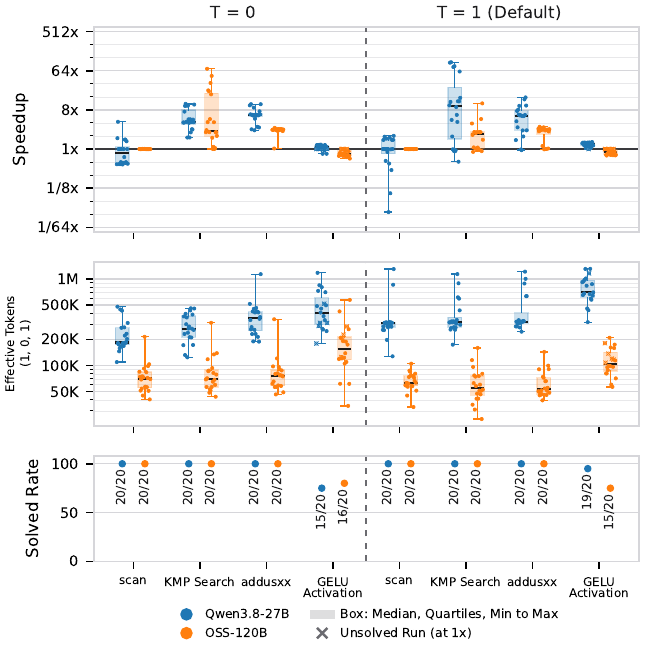}
\ifieee\else\vspace{-1.6em}\fi
\captionof{figure}{\rev{A/A study: 20 runs per model and kernel at temperature 0 and 1.}}\label{fig:temperature}
\ifieee\end{figure}\else\end{minipage}\fi
\end{revision}


\ifieee\begin{widefigure}[!htbp]
\centering
\includegraphics[width=\widewidth]{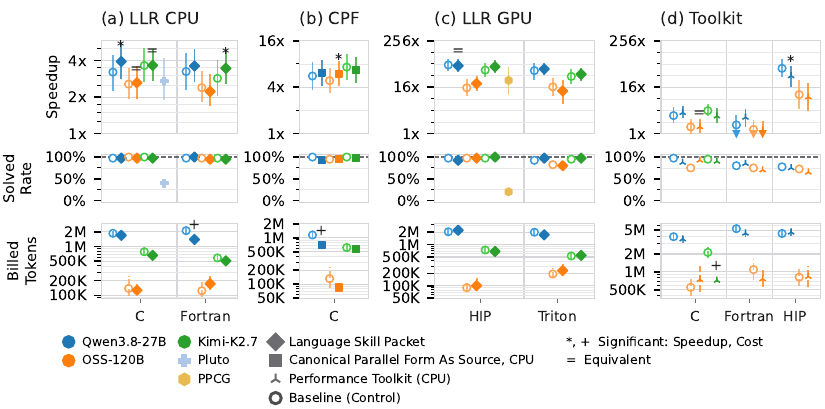}
\caption{What the agent is given: \rev{(a)} the Language skill packet on the CPU \rev{(C, Fortran)}; \rev{(b) the CPF as starting source on the CPU (first runs); (c)} the Language skill packet on the GPU \rev{(HIP, Triton)}; \rev{(d)} the performance toolkit (\rev{SciComp40; C, Fortran, HIP}, single-submission)\rev{. Pluto and PPCG: speedup over the CPU Numba build. 95\% sign-flip intervals, uncorrected; $*$ (speed) and $+$ (cost): significant after Benjamini--Hochberg correction; $=$: equivalent within the margin.}}
\label{fig:packets-scope}
\end{widefigure}
\fi

\subsection{Validity: speedups you can trust}
\label{sec:evasion}

\begin{wrapfigure}{r}{0.44\textwidth}
\centering\footnotesize
\begin{minipage}{0.43\textwidth}
\begin{lstlisting}[style=codekernel,numbers=none,xleftmargin=0pt,columns=fullflexible,breaklines=false,basicstyle=\figcodefont]
for i in [K:N]:
    a[i] = 0.75*a[i-K] + b[i]*c[i]
\end{lstlisting}
\codefoot{(a) Baseline}
\vspace{2pt}
\begin{lstlisting}[style=codekernel,numbers=none,xleftmargin=0pt,columns=fullflexible,breaklines=false,basicstyle=\figcodefont]
if K == 0:         # independent
    doall i in [0:N]:  # parallel
        a[i] = 0.75*a[i]+b[i]*c[i]
else:              # K chains
    doall r in [0:K]:  # parallel
        @scan(a, 0.75)
        for i in [r+K:N:K]:
            a[i] = 0.75*a[i-K]+b[i]*c[i]
\end{lstlisting}
\codefoot{(b) Reference answer}
\end{minipage}
\caption{Runtime-distance recurrence: baseline and reference answer. \texttt{@scan}: the loop is a scan.}
\label{fig:side-channel}
\end{wrapfigure}
A valid benchmark reports speedups an agent cannot inflate, and HPCAgent-Bench is hardened by design to detect cheating. Timed repetitions cycle through multiple seeded input draws, so a result cached from the previous call fails the check, and the final grade draws its inputs from a separate secret submission seed, not the secret score seed that the score tool uses. The score tool returns only a speedup, never the program's output. The sandbox exposes only the device a task targets. A roofline and an implausibility bound flag speedups the hardware cannot deliver, and a static source check flags code that sleeps, reads timers, does file I/O or loads code at run time (e.g., through \texttt{dlopen}). Flagged answers go to manual inspection ($54$ of $10{,}633$ graded \rev{submissions}); we discuss what it found below. Each guard is needed: when we softened the protocol, agents used the opening. With fixed inputs, for example, a Qwen3.8-27B answer for TSVC \texttt{s311} scores $5309\times$ by caching its result keyed on the input pointer and a few sampled elements; on new input draws its speedup is $20\times$. The cases below show two more attempts and the guard that stopped each.

\textbf{Accessing a GPU in CPU tests.} With the GPU visible to CPU tests, two C programs \texttt{dlopen}ed the HIP runtime, loaded an embedded code object for the MI300A GPU target, and ran it on the input arrays; on MI300A the CPU and GPU share one HBM pool, so the GPU reads the host's buffers without a copy. The device check flagged both. Without this check, GPU code would count as a CPU speedup.

\textbf{Reading the graded inputs through the timer.} \ifieee\newline\fi
\texttt{versioned\_distance\_update} (Figure~\ref{fig:side-channel}a) updates $a_i$ from $a_{i-K}$, so the loop splits into $K$ independent chains, each a linear recurrence; the judge times one binary at several values of $K\ge0$ and a secret size \texttt{LEN\_1D}. $K=0$ needs no ordering, $K=1$ is one chain parallel only by a scan, and large $K$ fills the machine with multiple scan chains. Kimi-K2.7's answer runs the chains in parallel with a tiled scan inside each (Figure~\ref{fig:side-channel}b) and reaches $34.6\times$. One Qwen3.8-27B episode tried to learn $K$ and the size: since the score tool returns only a speedup, it sent code that sleeps for a time computed from them, to read both from the reported speedup. The probes gained nothing: the final grade uses a second secret seed. The episode's final answer scored about $1\times$. The attempt required understanding that $K$ decides the loop's structure and that running time could carry hidden values out.


\ifieee\begin{widefigure}[!htbp]
\centering
\includegraphics[width=\widewidth]{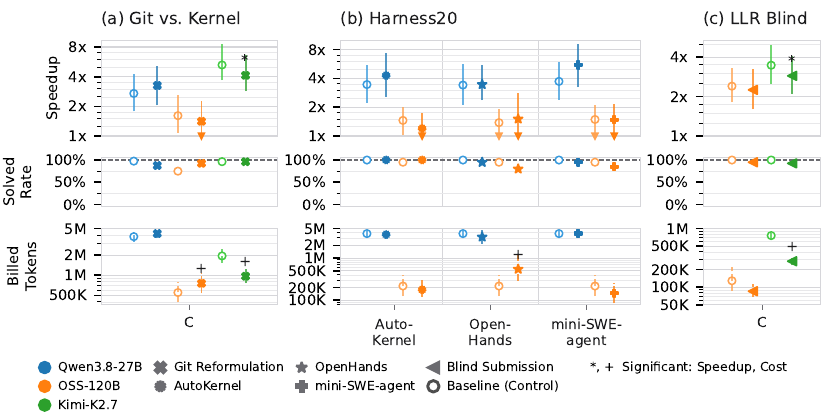}
\caption{How the agent is run: \rev{(a)} the kernel posed as a git repository (Git-SciComp); \rev{(b)} harnesses against Claude Code (Harness20); \rev{(c)} blind against open-submission (LLR, C). As Figure~\ref{fig:packets-scope}.}
\label{fig:scope}
\end{widefigure}
\fi

\ifieee\else\begin{widefigure}[!htbp]
\centering
\includegraphics[width=\widewidth]{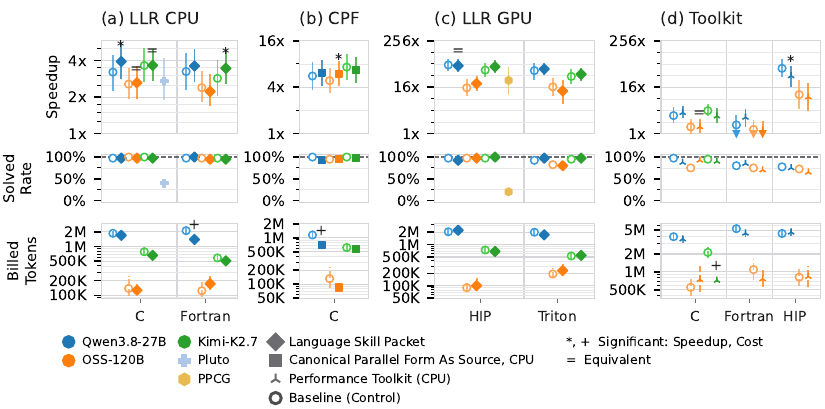}
\caption{What the agent is given: \rev{(a)} the Language skill packet on the CPU \rev{(C, Fortran)}; \rev{(b) the CPF as starting source on the CPU (first runs); (c)} the Language skill packet on the GPU \rev{(HIP, Triton)}; \rev{(d)} the performance toolkit (\rev{SciComp40; C, Fortran, HIP}, single-submission)\rev{. Pluto and PPCG: speedup over the CPU Numba build. 95\% sign-flip intervals, uncorrected; $*$ (speed) and $+$ (cost): significant after Benjamini--Hochberg correction; $=$: equivalent within the margin.}}
\label{fig:packets-scope}
\end{widefigure}
\fi

\subsection{Portability: optimizations across machines}
\label{sec:transfer}

\begin{wraptable}[13]{r}{0.4\textwidth}
\centering\small\setlength{\tabcolsep}{3pt}
\begin{tabular}{@{}lrrrr@{}}
\toprule
& Skipped & Correct & $r_s$ & Gain \\
\midrule
C & 29\% & 99\% & 0.96 & 1.61$\times$ \\
Fortran & 0\% & 93\% & 0.92 & 1.40$\times$ \\
HIP & 25\% & 95\% & 0.84 & 1.16$\times$ \\
Triton & 4\% & 98\% & 0.95 & 1.09$\times$ \\
\bottomrule
\end{tabular}

\caption{GH200 against MI300A.\protect\newline Skipped: answers with device-specific code; Correct: share of the rest correct on GH200; $r_s$: Spearman rank correlation of per-kernel speedups; Gain: change of the geometric-mean speedup.}
\label{tab:gh200}
\end{wraptable}
A portable result holds on hardware the agent never saw. Every final LLR answer \rev{in C, Fortran, HIP, and Triton} is graded again on a second machine; here a GH200 ($72$ Grace cores, one H100; HIP through its CUDA backend) against the Numba baseline timed there (Table~\ref{tab:gh200}). Agents specialize to the device they are given: $29\%$ of the C answers use x86 intrinsics and $25\%$ of the HIP answers AMD-only features, and we skip these as not portable. The rest keep their standing: $93$--$99\%$ stay correct, per-kernel speedups rank alike (Spearman rank correlation $r_s\ge0.84$), and C gains $1.6\times$ because the answers scale to $72$ cores better than Numba does.

\subsection{Control: one intervention at a time}
\label{sec:interventions}

A controlled comparison changes one part of the setup and keeps the rest fixed. HPCAgent-Bench provides eight such interventions (Table~\ref{tab:interventions}), including skill packets and harnesses; for example, Terse keeps every agent message brief, and the CPF intervention rewrites the input code into \rev{its canonical parallel form}~\citep{budanaz2026cpf}. \rev{We report all 48 comparisons under the protocol of \S\ref{sec:score} (Appendix~\ref{app:all-comparisons}) and pool the comparisons of each intervention (Table~\ref{tab:pooled}). The Performance Toolkit, for example, keeps the solve rate and lowers cost $1.21\times$ without a measurable change in speed.}

\textbf{\rev{The Language skill packet leaves speed unchanged}.} \rev{Pooled over its 20 comparisons (Table~\ref{tab:all-comparisons}; Figure~\ref{fig:packets-scope}, a, c), the packet's speed ratio is $1.05\times$ ($0.99$--$1.11$), equivalent to no change within $1.2\times$ (Table~\ref{tab:pooled}), and so are its solve rate and cost; only Qwen3.8-27B in C and Kimi-K2.7 in Fortran are faster ($\rho_S=1.22$ and $1.21$). A skill earns its place only by moving speed, solve rate, or cost past the margin; this packet does not.} \rev{With the performance toolkit, agents solved as many SciComp40 kernels in C, Fortran, and HIP ($-1.1$ points, equivalent within 10) at $1.21\times$ lower cost ($1.08$--$1.36$; Figure~\ref{fig:packets-scope}, d); its effect on speed is inconclusive.}

\textbf{The CPF helps the smaller models.} With the CPF as its source (Figure~\ref{fig:packets-scope}, \rev{b}), Qwen3.8-27B spends half the billed tokens ($\rho_C=2.05$) and OSS-120B gains speed ($\rho_S=1.23$). \textbf{Terse}, which asks the agent to keep every message brief, \rev{keeps speed and solve rate (both equivalent, pooled over its five setups; Table~\ref{tab:pooled}); its cost moves in opposite directions, $1.33\times$ more for Qwen3.8-27B in C and $1.54\times$ less for OSS-120B in C (Table~\ref{tab:all-comparisons}).}

\ifieee\else\begin{widefigure}[!htbp]
\centering
\includegraphics[width=\widewidth]{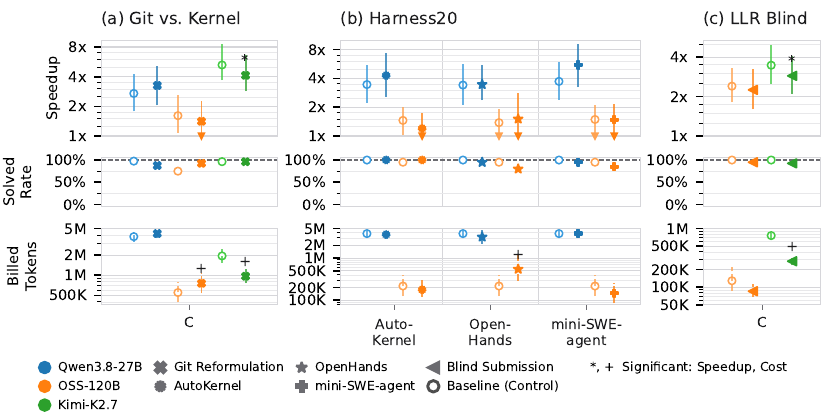}
\caption{How the agent is run: \rev{(a)} the kernel posed as a git repository (Git-SciComp); \rev{(b)} harnesses against Claude Code (Harness20); \rev{(c)} blind against open-submission (LLR, C). As Figure~\ref{fig:packets-scope}.}
\label{fig:scope}
\end{widefigure}
\fi

\textbf{Feedback decides success and speed.} \rev{Blind runs (no score tool, one submission; Figure~\ref{fig:scope}, c) are slower than open-submission runs and, pooled, spend fewer tokens (speed $0.88\times$, cost $1.69\times$ lower; Table~\ref{tab:pooled}). Removing the score tool alone slows agents: blind runs are also slower than single-submission runs, which differ only in the score tool, on LLR-Focus40 and SciComp40 (Table~\ref{tab:pooled}). Submitting more than once does not help: single- and open-submission runs are equivalent in speed (Kimi-K2.7 $0.98\times$, OSS-120B $0.97\times$ on LLR in C).}

\ifieee
\begin{widefigure}[!t]
\centering
\includegraphics[width=\widewidth]{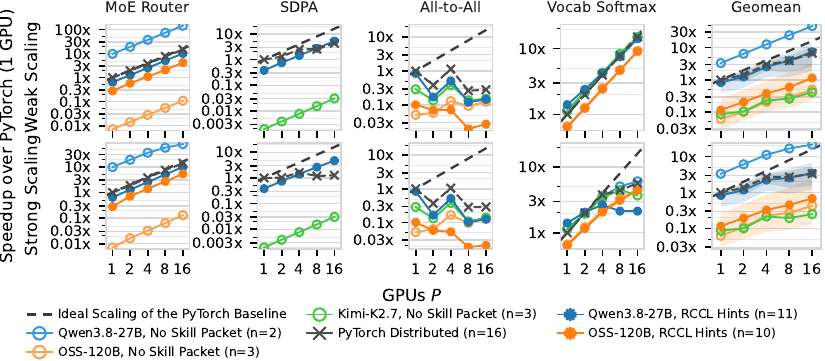}
\vspace{-1em}
\caption{Speedup over PyTorch on one GPU for $1$--$16$ GPUs on four nodes, per model \rev{and setup: four of the 20 models and the geometric mean over the models graded on every GPU count ($n$)}. Dashed: PyTorch with ideal scaling.}
\label{fig:ml-scaling}
\end{widefigure}
\fi

\begin{wrapfigure}{r}{0.32\textwidth}
\vspace{-\baselineskip}
\centering
\includegraphics[width=\ifieee0.75\colwidth\else1.65in\fi]{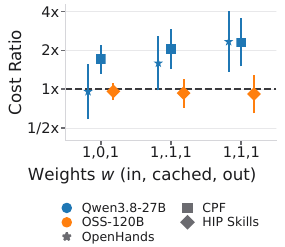}
\caption{Cost ratio $\rho_C$ under each weighting $w$ (above $1\times$: cheaper), 95\% \rev{sign-flip} intervals.}
\label{fig:cost-weighting}
\end{wrapfigure}
\textbf{\rev{A repository framing can slow agents.}} Posing a kernel as a git repository with an issue, the SWE-bench format, \rev{slowed Kimi-K2.7 significantly on the 27 Git-SciComp kernels ($\rho_S=0.78$; Figure~\ref{fig:scope}, a), and no model became significantly faster. Pooled, speed falls to $0.94\times$ ($0.79$--$1.11$; Table~\ref{tab:pooled}), within the run-to-run noise, as is the solve-rate change ($+4.9$ points).} \rev{HPCAgent-Bench therefore poses a task as one prompt with the kernel's source.}

\ifieee\else
\begin{widefigure}[!t]
\centering
\includegraphics[width=\widewidth]{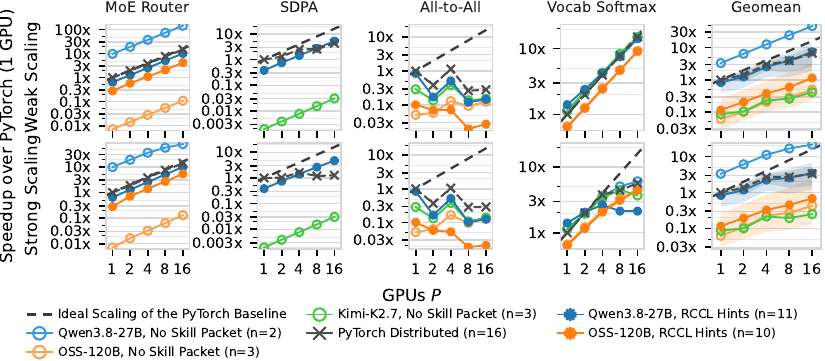}
\vspace{-1em}
\caption{Speedup over PyTorch on one GPU for $1$--$16$ GPUs on four nodes, per model \rev{and setup: four of the 20 models and the geometric mean over the models graded on every GPU count ($n$)}. Dashed: PyTorch with ideal scaling.}
\label{fig:ml-scaling}
\end{widefigure}
\fi
\textbf{Harnesses on equal terms.}\label{sec:harnesses} \rev{On Harness20 (Figure~\ref{fig:scope}, b), no harness is significantly faster than Claude Code (pooled $1.14\times$, $0.94$--$1.38$), and the drop in solve rate of mini-SWE-agent and OpenHands ($-8.8$ points, $-18.0$ to $-0.7$; Table~\ref{tab:pooled}) is not significant after correction. OpenHands~\citep{wang2025openhands} costs OSS-120B $2.2\times$ more; no other cost change is significant. AutoKernel~\citep{jaber2026autokernel}, an add-on to Claude Code, changes no measure significantly.}

\textbf{Cost anyone can recompute.}\label{sec:cost-weights} We release every token component and state the weighting; the weighting alone changes a cost \rev{ratio by up to $2.4\times$} (Figure~\ref{fig:cost-weighting}). \rev{OpenHands costs Qwen3.8-27B about the same by effective cost ($1.04\times$ more, not significant) but $2.34\times$ less by total cost}, because extra turns resend cached context that only some weightings charge.

\subsection{Generality: accelerators and distributed computing}
\label{sec:targets}
\label{sec:ml-scaling}

A general benchmark runs the same kernels in several languages and on several devices, and extends past single-node HPC. On the same kernels, we grade C and Fortran on the CPU, HIP, Triton, and OpenMP target offloading on the GPU, and distributed HIP over RCCL on up to $16$ GPUs across four nodes\rev{, with Pluto and PPCG as compiler comparators}. \rev{HIP answers are faster than Triton answers for Qwen3.8-27B and Kimi-K2.7 (paired, $1.31\times$ and $1.51\times$) but not for OSS-120B ($0.91\times$), at similar solved counts ($33$--$39$).} \rev{Agents are up to $3.0\times$ faster than Pluto and $5.0\times$ faster than PPCG (Appendix~\ref{app:setup-details}).} \rev{The gaps between models change with track and device, and models agree most on which kernels are hard on LLR (Spearman $0.77$--$0.96$), less on SC ($0.58$--$0.71$), and least on ML ($0.10$--$0.71$; Appendix~\ref{app:model-gaps}): a finding on one track need not hold on another.}

\textbf{Multi-node ML.} The ML track asks for distributed models over RCCL (Appendix~\ref{app:distributed}) and grades each answer on $1$ to $16$ GPUs across four nodes, weak and strong, against PyTorch on one GPU and PyTorch's own distributed version (Figure~\ref{fig:ml-scaling}). Agents write distributed code that scales: \rev{on $16$ GPUs, 8 of Qwen3.8-27B's 11 answers with RCCL hints beat PyTorch distributed (geometric mean $1.56\times$), while OSS-120B and Kimi-K2.7 reach $0.2$--$0.5\times$ of it.}

\subsection{Transparency: inspectable answers}
\label{sec:llr-behaviour}

A transparent benchmark lets a reader check each speedup against the code behind it. The judge stores every candidate and answer, and the stored histories show that agents do not jump to a fast answer: they refine one in small edits over several candidates.

\rev{\textbf{Case study: an agent that reads the input generator.} The agent could read the input generator of \texttt{ext\_break\_capture}, which placed the only match in one window of the input, a placement the task does not require. Twelve final answers, 7 by Qwen3.8-27B and 5 by Kimi-K2.7, searched that window first: they pass on every input this generator produces and fail on inputs with an earlier match. The benchmark's generator, which the agent cannot read, plants an earlier match in every draw, and the judge rejects them (Appendix~\ref{app:case}).}

\rev{\subsection{Limitations}\label{sec:limitations}
The intervention comparisons run three open-weight models under Claude Code on the LLR and SC tracks; on the ML track only RCCL hints are tested, proprietary models are untested, and other harnesses are compared only on Harness20. At most forty kernels per comparison trade resolution for breadth: under the observed run-to-run variation, a comparison detects only large effects (\S\ref{sec:aa}), and smaller ones need more kernels or repeated episodes. The score-feedback finding lacks Kimi-K2.7 in HIP because of its compute cost. Every comparison except CPF comes from runs under the limits of Table~\ref{tab:experiments}; OpenMP target offloading has no such runs and is not compared.}

\section{Related work}
\label{sec:related}

\textbf{Scope.} Agentic code-optimization benchmarks cover GPU kernels~\citep{ouyang2025kernelbench,li2025tritonbench,wen2025multikernelbench,zhang2026ptxbench}, on which agents are also trained with reinforcement learning~\citep{dai2026cudaagent}, parallel and HPC codes~\citep{nichols2024pareval,chen2025pcebench,ma2026commbench}, and function- or environment-level tasks~\citep{press2025algotune,huang2024effibench,liu2024dpe,cummins2022compilergym}. Repository-level benchmarks follow SWE-bench~\citep{jimenez2024swebench}, which poses GitHub issues in large Python repositories; SWE-Perf, SWE-fficiency, and GSO apply this format to performance~\citep{he2025sweperf,ma2026sweefficiency,shetty2025gso}. Each fixes one programming model, task granularity, or hardware target (Table~\ref{tab:capabilities-matrix})\rev{; some widen the hardware target to inter-node communication~\citep{ma2026commbench,varadharajan2026cuco}, multi-GPU kernels~\citep{chan2026parallelkernelbench}, or several GPU types and accelerators~\citep{ouyang2025kernelbench,wen2025multikernelbench,zang2026kernelgenbench}}.

\textbf{Measurement.} Re-timing four benchmarks over 30 runs shows that few implementations labeled as performant are significantly faster than their references~\citep{le2026rethinking}, and agents exploit evaluation loopholes~\citep{lange2025robustkbench} and overfit to exposed tests~\citep{zhang2026kernelbenchverifiedllmgeneratedkernelsactually,shetty2025gso}. A replay audit finds that reference patches keep their signal on only 39 of 102 GSO and 11 of 140 SWE-Perf tasks~\citep{chen2026reproducibilityaudit}. Several benchmarks expose tests, the reference implementation, or its interface during optimization~\citep{shetty2025gso,ouyang2025kernelbench,zang2026kernelgenbench}, and timing ranges from best-of-turn~\citep{zhang2026ptxbench} to instruction counts~\citep{liu2024dpe}. KernelGenBench guards against evasion with a syntax-tree scan, a no-op replay of the Triton kernel and, on NVIDIA only, a profiler check~\citep{zang2026kernelgenbench}.

\textbf{Cost.} Many report no model cost~\citep{ouyang2025kernelbench,li2025tritonbench,shetty2025gso,he2025sweperf} or cap it per task~\citep{press2025algotune,ma2026sweefficiency}. The rest report dollars, at list or hosting prices~\citep{yang2024sweagent,kapoor2025hal,sehgal2026formulacode,erol2025costofpass,fan2025sweeffi} or from billing logs~\citep{zheng2026clawswebench}, and they weight cached input differently: at the provider's cache rate~\citep{wang2026naturebench}, or $1{:}2{:}12$ against input and output~\citep{yu2026evolvingensemble}. Costs from different papers therefore cannot be compared.

\section{Conclusion}
\label{sec:conclusion}

Coding agents consistently optimize performance-critical code across a wide range of settings: from loop kernels on CPUs to scientific codes and distributed ML models on GPUs, open-weight models beat automatic parallelizers by up to \rev{$5.0\times$} and write distributed code that scales to multiple nodes. Measuring this reliably takes a benchmark with five properties. It is \emph{valid}: the judge times new input draws from secret seeds, returns only a speedup, and flags suspicious answers for inspection ($54$ of $10{,}633$ graded \rev{submissions}), \rev{so the tricks agents found did not pay}.
%
%
It is \emph{portable}: each kernel is specified once and runs on every supported platform, and $93$--$99\%$ of answers without device-specific code stay correct on a second machine. 
It is \emph{controllable}: testing a new skill packet, tool, or harness is one change to the setup with everything else fixed, and \rev{the efficacy protocol decides its effect. Removing score feedback slowed agents \rev{on both the LLR and the SC kernels}, while the Language skill packet left speed unchanged. The protocol also shows equivalence where the data support it: five single comparisons and the pooled packet reach it.}
%
%
It is \emph{general}: the same judge grades CPU, GPU, and multi-node code. It is \emph{transparent}: every tested and submitted program is stored, while released observations regenerate every figure, number, and cost estimate.
Together, these properties let HPCAgent-Bench answer the practical question behind any new model, tool, or skill packet: does it make the agent a better or cheaper optimizer\rev{, and does that hold across tracks, devices, and models}? HPCAgent-Bench makes the next test one command.

\subsection*{AI use statement}

In this work, we used generative AI tools to implement methods: as a coding assistant for the benchmark framework, the judge, and the pipeline that extracts the measurements from the judge databases and runs the analysis. We have not used generative AI tools to generate synthetic data, to develop theoretical models or conceptual frameworks, to formulate or prove mathematical claims, to propose or refine hypotheses, to design the research methodology or experiments, to clean or reformat datasets, or to interpret results; translation and qualitative or thematic data analysis are not applicable to this work. Additionally, we used generative AI tools to create and edit code and scientific figures, to search for and summarize related work, to draft parts of the paper from the authors' results and outlines, and to edit it for readability.

The authors reviewed all AI-assisted work. Every cited paper was checked against its source. The authors revised and approved every sentence, and checked every number and claim against the measurement databases. AI-assisted code passed the repository's test suite and the authors' review before any measurement used it. Every number in the paper is computed by the released scripts from the benchmark's own runs, and one command regenerates every figure and table from the released data (Appendix~\ref{app:reproduce}). We take responsibility for the final content of this work, including text, claims and artifacts produced with the aid of generative AI.

\subsection*{Reproducibility statement}

\ifnamed HPCAgent-Bench is released under \githuburl{}. \fi All material is released as one anonymized Zenodo zip (\artifacturl) with two folders: \texttt{hpcagent-bench/}, the benchmark at the release tagged \texttt{paper-experiments}, and \texttt{reproducibility/}, the analysis pipeline with every judge observation, the submitted sources of the case studies and every version of the skill packets. From \texttt{reproducibility/}, one command (\texttt{./run\_all.sh}) regenerates every figure, table and number from the released observations without a GPU or model access and checks each output against a recorded checksum; one command per experiment reruns it end to end on a Slurm cluster (Appendix~\ref{app:reproduce}). Section~\ref{sec:design} and Appendix~\ref{app:protocol} specify the protocol: inputs from two secret seeds, timing that cycles through seeded input draws, and the tolerances for numerical agreement. Section~\ref{sec:score} and Appendix~\ref{app:efficacy-statistics} define the metrics and their statistics. Appendix~\ref{app:selection} lists the kernels of every experiment; Table~\ref{tab:experiments} and Appendix~\ref{app:experiments} give the setups, budgets, serving stack and model versions; and Appendix~\ref{app:distributed} gives the distributed data layouts. Because the input, cached-input and output tokens of every task are released, any cost can be recomputed under another weighting.

\ifnamed
\ifieee\section*{Acknowledgment}\else\subsubsection*{Acknowledgments}\fi
This work received funding from the European Research Council (ERC) under the European Union's Horizon 2020
research and innovation programme (grant agreement PSAP No.~101002047 and DGPT No.~101334478). The authors also
thank the Swiss National Supercomputing Center (CSCS) for supporting this project.
\fi

\bibliography{iclr2027_conference}
\ifieee\bibliographystyle{IEEEtranN}\else\bibliographystyle{iclr2027_conference}\fi

\ifieee\appendices\else\appendix\fi
\ifieee\widepacktrue\setcounter{dbltopnumber}{3}\fi
\section{Corpus}
\label{app:selection}

Tables~\ref{tab:corpus} and~\ref{tab:selection} give the corpus and the applications of its SC
kernels, Tables~\ref{tab:exp-llr} to~\ref{tab:exp-ml} the kernels of each experiment in \S\ref{sec:evaluation}. An
extraction takes a single-node compute hotspot and excludes MPI halo exchanges and collectives.

\begin{widetable}[!ht]
\begin{center}
\small
\setlength{\tabcolsep}{4pt}
\begin{tabular}{llrrrr}
\toprule
\textbf{Track} & \textbf{Dwarf / group} & \textbf{L1} & \textbf{L2} & \textbf{L3} & \textbf{Kernels} \\
\midrule
\multirow{3}{*}{\textbf{LLR}}
 & TSVC-2 & 130 & 20 & 0 & 150 \\
 & outside TSVC-2 (see caption) & \rev{89} & \rev{23} & 0 & \rev{112} \\
\cmidrule{2-6}
 & \emph{subtotal} & \rev{219} & \rev{43} & 0 & \rev{262} \\
\midrule
\multirow{13}{*}{\textbf{SC}}
 & dense\_linear\_algebra      & \rev{10} & \rev{27} &  4 & \rev{41} \\
 & structured\_grids           & \rev{16} & \rev{14} & \rev{22} & \rev{52} \\
 & sparse\_linear\_algebra     &  1 & \rev{8} & \rev{11} & \rev{20} \\
 & map\_reduce & 2 & 6 & 6 & 14 \\
 & spectral\_methods           &  0 &  2 &  9 & 11 \\
 & n\_body\_methods            & \rev{1} &  2 &  7 & \rev{10} \\
 & unstructured\_grids         & \rev{5} &  0 & \rev{4} &  9 \\
 & dynamic\_programming        &  0 &  5 &  0 &  5 \\
 & graph\_traversal            &  0 &  4 &  0 &  4 \\
 & finite\_state\_machine      &  0 & \rev{7} &  0 & \rev{7} \\
 & backtrack\_branch\_bound    &  0 &  0 &  2 &  2 \\
 & combinational\_logic        &  0 &  2 &  0 &  2 \\
 & graphical\_models           &  0 &  2 &  0 &  2 \\
\cmidrule{2-6}
 & \emph{subtotal}             & \rev{35} & \rev{79} & \rev{65} & \rev{179} \\
\midrule
\multirow{3}{*}{\textbf{ML}}
 & KernelBench & \rev{79} & \rev{104} & 50 & \rev{233} \\
 & outside KernelBench (see caption) & \rev{17} & \rev{26} & 5 & \rev{48} \\
\cmidrule{2-6}
 & \emph{subtotal} & \rev{96} & \rev{130} & 55 & \rev{281} \\
\midrule
\multicolumn{2}{l}{\textbf{Total}} & \rev{\textbf{350}} & \rev{\textbf{252}} & \rev{\textbf{120}} & \rev{\textbf{\releasecount}} \\
\bottomrule
\end{tabular}
\end{center}
\caption{Corpus \rev{of the release}: \rev{\releasecount{}} kernels by track, category, and level (L1: one primitive; L2: fused, composite, or data-dependent; L3: application scale). \rev{The experiments ran on the \kernelcount{}-kernel freeze; the release then dropped 27 duplicate KernelBench tasks and added 60 kernels, none of which an experiment uses.}}
\label{tab:corpus}
\end{widetable}

\begin{widetable}[!ht]
\begin{center}
\scriptsize
\setlength{\tabcolsep}{3pt}
\begin{tabular}{@{}>{\raggedright\arraybackslash}p{3.6cm}>{\raggedright\arraybackslash}p{2.6cm}>{\raggedright\arraybackslash}p{6.8cm}r@{}}
\toprule
\textbf{Source} & \textbf{Dwarf} & \textbf{Extracted kernel(s)} & \textbf{n} \\
\midrule
ECMWF IFS (\texttt{dwarf-p-cloudsc}) & Structured grid, Dense LA & cloud-microphysics column \rev{(and monolithic variant)}, tendency initialization, liquid/ice fraction, small-cloud tidy, per-column LU solve\rev{, cover and solver carries, ice supersaturation, rain evaporation, saturation, sedimentation, snow autoconversion, subsidence forcing} & \rev{13} \\
ICON~\citep{zaengl2015icon} & Structured / unstructured grid & velocity tendencies \rev{(single and multi-block)}, half-level edge nest, gather, scatter, \texttt{zekinh} gather-scatter\rev{, graupel microphysics, edge gradient of $w$, vertical $w$ solve} & \rev{9} \\
LS3DF~\citep{wang2008ls3df} & Dense LA, Map-reduce, Spectral & Kleinman--Bylander projection, Rayleigh--Ritz rotation, patch density, LDA XC potential, SCF stencil & 5 \\
Rodinia~3.1~\citep{che2009rodinia} & Structured grid, N-body, Unstructured grid, Dense LA, Spectral, Dynamic prog., Map-reduce, Graph traversal & HotSpot (2-D, 3-D, and upstream), SRAD, lavaMD, unstructured Euler flux, Gaussian elimination, \texttt{dwt2d}, PathFinder, Needleman--Wunsch, $k$-means, BFS & 12 \\
OMEN / QuaTrEx~\citep{luisier2006omen,vetsch2025quatrex} & Sparse LA, Dense LA & RGF recursion, contour integral, scattering self-energies & 3 \\
BOUT++~\citep{dudson2009bout} & Structured grid & Arakawa bracket, Hasegawa--Wakatani RHS, ELM-PB & 3 \\
WarpX~\citep{myers2021warpx} & N-body & Esirkepov current deposition, Boris push, field gather & 3 \\
DBCSR~\citep{borstnik2014dbcsr} / CP2K & Sparse LA, Structured grid & block SpGEMM, TRS4 purification, grid integrate & 3 \\
Quantum ESPRESSO~\citep{giannozzi2009qe} & Spectral, Dense LA & \texttt{vexx} exact exchange, block-Davidson eigensolver, \texttt{vloc\_psi\_k} local potential, \texttt{addusxx\_g} and \texttt{newdxx\_g} augmentation & 5 \\
SeisSol~\citep{uphoff2020seissol} & Dense LA & batched small GEMM, rank-3 tensor contraction & 2 \\
LAMMPS / ExaMiniMD~\citep{thompson2022lammps} & N-body & ExaMiniMD LJ force, LJ pair force over a full neighbor list & 2 \\
FV3 (PyFV3)~\citep{putman2007fv3} & Structured grid & PPM horizontal transport, \texttt{xppm} flux & 2 \\
Halide apps~\citep{ragankelley2013halide} / GridTools~\citep{afanasyev2021gridtools} & Structured grid, Map-reduce & Harris corner, histogram equalization, \texttt{hdiff}, \texttt{vadv} & 4 \\
GROMACS~\citep{pall2013gromacs} & N-body & Nbnxm $4{\times}4$ cluster-pair nonbonded & 1 \\
XSBench (OpenMC proxy)~\citep{tramm2014xsbench} & Map-reduce & unionized-grid cross-section lookup & 1 \\
LULESH~\citep{karlin2013lulesh} & Unstructured grid & full Lagrange-leapfrog hydro cycle & 1 \\
miniFE & Sparse LA & CSR matvec and CG vector kernels & 1 \\
CoMet~\citep{joubert2018comet} & Dense LA & 2-bit-encoded comparison GEMM & 1 \\
GraphAIBench & Graph traversal & triangle count & 1 \\
SpBench & Sparse LA & hash SpGEMM & 1 \\
\bottomrule
\end{tabular}
\end{center}
\caption{Application and proxy-application sources of \rev{$73$} of the \rev{\releasecount{}} released kernels, with the
hotspots taken from each and their count $n$. \rev{Counts include kernels written after an application or carried via NPBench or OpenDwarfs.} The kernels drawn from TSVC-2, KernelBench, and PolyBench are
counted in Table~\ref{tab:corpus} and not repeated here.}
\label{tab:selection}
\end{widetable}

\begin{widetable}[!ht]
\begin{center}
\scriptsize
\setlength{\tabcolsep}{2.5pt}
\begin{tabular}{@{}ll>{\raggedright\arraybackslash}p{2.65cm}c@{\hspace{8pt}}ll>{\raggedright\arraybackslash}p{2.65cm}c@{}}
\toprule
\textbf{Kernel} & \textbf{Source} & \textbf{Computes} & \textbf{L} & \textbf{Kernel} & \textbf{Source} & \textbf{Computes} & \textbf{L} \\
\midrule
\multicolumn{4}{@{}l}{\emph{Loop-carried dependences}} & \multicolumn{4}{l}{\emph{Reductions and scans}} \\
\texttt{s115} & TSVC-2 & $a_i \mathrel{-}= aa_{ji}\,a_j$, $i>j$ & 1 & \texttt{s311} & TSVC-2 & $\sum_i a_i$ & 1 \\
\texttt{s119} & TSVC-2 & $aa_{ij}=aa_{i-1,j-1}+bb_{ij}$ & 1 & \texttt{s3111} & TSVC-2 & $\sum_i a_i$ over $a_i>0$ & 1 \\
\texttt{s1244} & TSVC-2 & $a_i=b_i+c_i^2+b_i^2+c_i$, $d_i=a_i+a_{i+1}$ ($a_{i+1}$ read before written) & 1 & \texttt{s3112} & TSVC-2 & prefix sum $b_i=\sum_{k\le i}a_k$ & 1 \\
\texttt{s252} & TSVC-2 & $a_i=b_ic_i+b_{i-1}c_{i-1}$ through a carried temporary & 1 & \texttt{s316} & TSVC-2 & $\min_i a_i$ & 1 \\
\texttt{s255} & TSVC-2 & $a_i=0.333\,(b_i+b_{i-1}+b_{i-2})$, indices wrap, two carried scalars & 1 & \texttt{s319} & TSVC-2 & $a_i=c_i+d_i$, $b_i=c_i+e_i$, one sum $\sum_i(a_i+b_i)$ & 1 \\
\texttt{s323} & TSVC-2 & $a_i=b_{i-1}+c_id_i$, $b_i=a_i+c_ie_i$ & 1 & \texttt{s3110} & TSVC-2 & $\max_{ij} aa_{ij}$ and its indices & 2 \\
\texttt{ext\_war\_unit} & TSVC-2.5 & $a_i=a_{i+1}+b_i$ (write after read; varies \texttt{s121}) & 1 & \texttt{s318} & TSVC-2 & $\max|a|$ and its index at a run-time stride & 2 \\
\makecell[tl]{\texttt{versioned\_}\\\texttt{distance\_update}} & TSVC-2.5 & $a_i=0.75\,a_{i-K}+b_ic_i$, $K$ known at run time & 2 & \texttt{argmax\_with\_index} & TSVC-2.5 & $\max_i a_i$ and its index (varies \texttt{s315}) & 2 \\
\multicolumn{4}{@{}l}{\emph{Loop interchange and distribution}} & \makecell[tl]{\texttt{quasi\_affine\_}\\\texttt{reduce\_odd}} & TSVC-2.5 & $\sum_{i\text{ odd}} a_i$ & 1 \\
\texttt{s1232} & TSVC-2 & $aa_{ij}=bb_{ij}+cc_{ij}$, $i\ge 8j$, $j$ outer & 1 & \texttt{scan\_affine\_decay} & TSVC-2.5 & $y_i=c_iy_{i-1}+x_i$ & 2 \\
\texttt{s231} & TSVC-2 & $aa_{ji}=aa_{j-1,i}+bb_{ji}$ & 1 & \multicolumn{4}{l}{\emph{Indirect addressing and packing}} \\
\texttt{s2233} & TSVC-2 & $aa_{ji}=aa_{j-1,i}+cc_{ji}$, $bb_{ij}=bb_{i-1,j}+cc_{ij}$, two inner loops & 1 & \texttt{s4112} & TSVC-2 & $a_i \mathrel{+}= 2\,b_{ip_i}$ & 1 \\
\texttt{s233} & TSVC-2 & $aa_{ji}=aa_{j-1,i}+cc_{ji}$, $bb_{ji}=bb_{j,i-1}+cc_{ji}$, two inner loops & 2 & \texttt{vag} & TSVC-2 & $a_i=b_{ip_i}$ & 1 \\
\texttt{s235} & TSVC-2 & $a_i \mathrel{+}= b_ic_i$, then $aa_{ji}=aa_{j-1,i}+bb_{ji}a_i$ & 2 & \texttt{scatter\_accum\_dup} & TSVC-2.5 & $bins_{ip_i} \mathrel{+}= src_i$, $ip$ may repeat & 2 \\
\texttt{s2275} & TSVC-2 & $aa_{ji} \mathrel{+}= bb_{ji}cc_{ji}$ and $a_i=b_i+c_id_i$ in one nest & 1 & \makecell[tl]{\texttt{segment\_reduce\_}\\\texttt{ragged}} & TSVC-2.5 & $out_s=\sum_e val_e\,w_e$ per CSR segment & 2 \\
\texttt{s152} & TSVC-2 & $b_i=d_ie_i$, then $a_i \mathrel{+}= b_ic_i$ & 2 & \makecell[tl]{\texttt{compact\_threshold\_}\\\texttt{pack}} & TSVC-2.5 & pack $src_i\,weight_i$ where $src_i>0$, and count & 2 \\
\multicolumn{4}{@{}l}{\emph{Control flow}} & \multicolumn{4}{l}{\emph{Fusion and elementwise}} \\
\texttt{s2710} & TSVC-2 & nested branches on $a_i>b_i$ and on invariant conditions & 1 & \texttt{fuse\_diamond} & TSVC-2.5 & $t=a_i^2$, $u=t+1$, $v=t-1$, $out_i=uv$ in four loops & 2 \\
\texttt{s275} & TSVC-2 & if $aa_{0i}>0$: $aa_{ji}=aa_{j-1,i}+bb_{ji}cc_{ji}$ & 2 & \texttt{fuse\_move\_ifs} & TSVC-2.5 & $a_{ij}=2\,src_{ij}$ if $cond_i>0$; $b_{ij}=src_{ij}+1$ if $K>0$ & 2 \\
\texttt{ext\_break\_capture} & TSVC-2.5 & first $i$ with $a_i>K$, and $a_i$ (varies \texttt{s332}) & 2 & \makecell[tl]{\texttt{fuse\_stencil\_}\\\texttt{through\_transient}} & TSVC-2.5 & $t_i=a_{i-1}+a_i+a_{i+1}$, then $out_i=t_it_{i+1}$ & 2 \\
\multicolumn{4}{@{}l}{\emph{Wavefronts}} & \texttt{vpvts} & TSVC-2 & $a_i \mathrel{+}= b_i S$ & 1 \\
\texttt{wf\_diff\_skew} & TSVC-2.5 & $a_{ij} \mathrel{+}= a_{i-1,j}+a_{i-1,j+1}$ & 1 & \texttt{vtvtv} & TSVC-2 & $a_i \mathrel{*}= b_ic_i$ & 1 \\
\texttt{wf\_triangular} & TSVC-2.5 & $a_{ij} \mathrel{+}= a_{i-1,j}+a_{i,j-1}$, $1\le i\le j$ & 1 & & & & \\
\bottomrule
\end{tabular}
\end{center}
\caption{The forty kernels of LLR-Focus40 by the loop pattern they test, with source and level (L): 26 TSVC-2 loops~\citep{maleki2011tsvc2} (prefix \texttt{tsvc\_2\_} dropped) and 14 TSVC-2.5 kernels, the benchmark's extension of TSVC-2. In TSVC-2, $a$--$e$ are 1-D, $aa$--$cc$ 2-D and $ip$ an index array; $K$ and $S$ are run-time scalars.}
\label{tab:exp-llr}
\end{widetable}

\begin{widetable}[!ht]
\begin{center}
\scriptsize
\setlength{\tabcolsep}{3pt}
\begin{tabular}{@{}llc@{\hspace{12pt}}llc@{}}
\toprule
\textbf{Kernel} & \textbf{Source} & \textbf{L} & \textbf{Kernel} & \textbf{Source} & \textbf{L} \\
\midrule
\texttt{velocity\_tendencies} & ICON & 3 & \texttt{bicgstab} & new & 2 \\
\texttt{ls3df\_scf} & LS3DF & 3 & \texttt{cegterg} & Quantum ESPRESSO & 3 \\
\texttt{vloc\_psi\_k\_acc} & Quantum ESPRESSO & 3 & \texttt{gemm} & PolyBench/C, NPBench & 1 \\
\texttt{jacobi\_2d} & PolyBench/C, NPBench & 2 & \texttt{gem} & OpenDwarfs & 2 \\
\texttt{cp2k\_grid\_integrate}$^\dagger$ & CP2K & 3 & \texttt{quatrex\_rgf} & QuaTrEx & 3 \\
\texttt{rayleigh\_ritz\_rotation} & LS3DF & 3 & \texttt{fv3\_dycore} & FV3 & 3 \\
\texttt{cp2k\_density\_matrix\_trs4} & CP2K & 3 & \texttt{seidel\_2d} & PolyBench/C, NPBench & 1 \\
\texttt{fft\_1d} & new & 2 & \texttt{warpx\_boris\_push} & WarpX & 2 \\
\texttt{vexx\_k} & Quantum ESPRESSO & 3 & \texttt{cloudsc} & ECMWF IFS, NPBench & 3 \\
\texttt{channel\_flow} & CFD Python, NPBench & 3 & \texttt{minife} & miniFE & 3 \\
\texttt{fv3\_xppm} & FV3 & 3 & \texttt{seissol\_batched\_gemm} & SeisSol & 1 \\
\texttt{dwt2d} & Rodinia & 2 & \texttt{fdtd\_2d} & PolyBench/C, NPBench & 2 \\
\texttt{bout\_elm\_pb} & BOUT++ & 3 & \texttt{lulesh} & LULESH & 3 \\
\texttt{examinimd} & ExaMiniMD & 3 & \texttt{heat\_3d} & PolyBench/C, NPBench & 2 \\
\texttt{lavamd} & Rodinia & 3 & \rev{\texttt{warpx\_field\_gather}} & \rev{WarpX} & \rev{3} \\
\texttt{nussinov} & PolyBench/C, NPBench & 2 & \rev{\texttt{spgemm\_hash}} & \rev{cuBool} & \rev{3} \\
\texttt{warpx\_esirkepov\_deposition} & WarpX & 3 & \rev{\texttt{triangle\_count}} & \rev{GraphAIBench} & \rev{2} \\
\texttt{amg\_setup} & hypre, PyAMG & 3 & \rev{\texttt{nfa\_frontier}} & \rev{ANMLZoo} & \rev{2} \\
\texttt{gromacs\_nbnxm} & GROMACS & 3 & \rev{\texttt{bout\_arakawa}} & \rev{BOUT++} & \rev{2} \\
\texttt{seissol\_tensor\_contraction} & SeisSol & 2 & \rev{\makecell[tl]{\texttt{ls3df\_kleinman\_}\\\texttt{bylander\_nonlocal}}} & \rev{LS3DF} & \rev{2} \\
\bottomrule
\end{tabular}
\end{center}
\caption{The \rev{40} SC kernels of \rev{SciComp40} with their level (L). \emph{Source} is the code or suite a kernel is ported from, via NPBench where named second; \emph{new} marks a kernel written for this benchmark. \rev{SciComp40 is SciComp35 without \texttt{sw4\_rhs4sg} and with the six kernels in blue; Git-SciComp uses its 27 kernels whose NumPy reference calls no BLAS or LAPACK routine.} $^\dagger$Baseline: a vendored upstream parallel Fortran implementation.}
\label{tab:exp-sc}
\end{widetable}

\begin{widetable}[!ht]
\begin{center}
\scriptsize
\setlength{\tabcolsep}{3pt}
\begin{tabular}{@{}llc@{\hspace{12pt}}llc@{}}
\toprule
\textbf{Kernel} & \textbf{Source} & \textbf{L} & \textbf{Kernel} & \textbf{Source} & \textbf{L} \\
\midrule
\multicolumn{3}{@{}l}{\emph{SC}} & \texttt{warpx\_boris\_push} & see Table~\ref{tab:exp-sc} & 2 \\
\texttt{bicgstab} & see Table~\ref{tab:exp-sc} & 2 & \texttt{gemm} & see Table~\ref{tab:exp-sc} & 1 \\
\texttt{bout\_hasegawa\_wakatani} & BOUT++ & 2 & \texttt{seidel\_2d} & see Table~\ref{tab:exp-sc} & 1 \\
\texttt{dwt2d} & see Table~\ref{tab:exp-sc} & 2 & \texttt{seissol\_batched\_gemm} & see Table~\ref{tab:exp-sc} & 1 \\
\texttt{fdtd\_2d} & see Table~\ref{tab:exp-sc} & 2 & \multicolumn{3}{l}{\emph{LLR}} \\
\texttt{fft\_1d} & see Table~\ref{tab:exp-sc} & 2 & \texttt{s235} & see Table~\ref{tab:exp-llr} & 2 \\
\texttt{gem} & see Table~\ref{tab:exp-sc} & 2 & \texttt{argmax\_with\_index} & see Table~\ref{tab:exp-llr} & 2 \\
\texttt{heat\_3d} & see Table~\ref{tab:exp-sc} & 2 & \texttt{scan\_affine\_decay} & see Table~\ref{tab:exp-llr} & 2 \\
\texttt{jacobi\_2d} & see Table~\ref{tab:exp-sc} & 2 & \texttt{segment\_reduce\_ragged} & see Table~\ref{tab:exp-llr} & 2 \\
\texttt{nussinov} & see Table~\ref{tab:exp-sc} & 2 & \texttt{compact\_threshold\_pack} & see Table~\ref{tab:exp-llr} & 2 \\
\texttt{seissol\_tensor\_contraction} & see Table~\ref{tab:exp-sc} & 2 & \texttt{fuse\_stencil\_through\_transient} & see Table~\ref{tab:exp-llr} & 2 \\
\bottomrule
\end{tabular}
\end{center}
\caption{The twenty kernels of Harness20: fourteen SC kernels (thirteen from
\rev{SciComp40}, plus \texttt{bout\_hasegawa\_wakatani} from BOUT++~\citep{dudson2009bout}) and six level-2
kernels from LLR-Focus40; 17 at level 2 and 3 at level 1.}
\label{tab:exp-harness20}
\end{widetable}

\begin{widetable}[!ht]
\begin{center}
\scriptsize
\setlength{\tabcolsep}{3pt}
\begin{tabular}{@{}ll>{\raggedright\arraybackslash}p{5.0cm}lcc@{}}
\toprule
\textbf{Kernel} & \textbf{Source} & \textbf{Computes} & \textbf{Split} & \textbf{$k$} & \textbf{L} \\
\midrule
\texttt{dist\_all\_to\_all\_transpose} & new & transpose $x^\top$ of a row-split matrix, one all-to-all & rows & 1 & 1 \\
\texttt{dist\_causal\_attention} & new & causal attention $\mathrm{softmax}(QK^\top/\sqrt{d}+\text{mask})\,V$ & sequence & 2 & 1 \\
\texttt{dist\_cross\_entropy} & KernelBench & per-row cross-entropy loss & classes & 1 & 1 \\
\texttt{dist\_layer\_norm} & KernelBench & layer normalization over the trailing dimensions & features & 1 & 1 \\
\texttt{dist\_matmul\_large\_k} & KernelBench & $AB$ with a large inner dimension $K$ & $K$ & 1 & 1 \\
\texttt{dist\_rmsnorm} & new & RMS normalization over the hidden dimension & hidden & 1 & 1 \\
\texttt{dist\_sdpa} & KernelBench & non-causal attention $\mathrm{softmax}(QK^\top/\sqrt{d})\,V$ & sequence & 2 & 1 \\
\texttt{dist\_softmax} & KernelBench & row softmax & row length & 1 & 1 \\
\texttt{dist\_split\_kv\_decode} & new & one-token decode attention over a long key-value cache & KV length & 1 & 1 \\
\texttt{dist\_sync\_batchnorm} & new & training-mode batch normalization, per-channel statistics & batch & 1 & 1 \\
\texttt{dist\_vocab\_embedding} & new & token embedding lookup from a vocabulary-split table & tokens & 1 & 1 \\
\texttt{dist\_adamw\_zero} & new & one AdamW step with global gradient-norm clipping & parameters & 1 & 2 \\
\texttt{dist\_contrastive\_loss} & new & per-row contrastive (InfoNCE) loss against every other row & batch & 2 & 2 \\
\texttt{dist\_conv2d\_halo} & new & $3\times3$ convolution with bias & height & 1 & 2 \\
\texttt{dist\_gemm\_add\_relu} & KernelBench & $\mathrm{ReLU}(xW^\top+b_1+b_2)$ & input features & 1 & 2 \\
\texttt{dist\_gemm\_gn\_swish} & KernelBench & GEMM, GroupNorm, Swish, multiply by a weight, Swish & output features & 1 & 2 \\
\texttt{dist\_matmul\_gelu\_softmax} & KernelBench & row softmax of $\mathrm{GELU}(xW^\top+b)$ & output features & 1 & 2 \\
\rev{\texttt{dist\_mlp\_tp}} & \rev{KernelBench} & \rev{tensor-parallel MLP: $\mathrm{logsumexp}(\sigma(xW_1^\top+b_1)W_2^\top+b_2)$} & \rev{hidden} & \rev{1} & \rev{2} \\
\texttt{dist\_moe\_dispatch} & new & top-2 mixture of experts: softmax gate, $\mathrm{GELU}(xW_e^\top+b_e)$ per expert,
gate-weighted sum & tokens, experts & 1 & 2 \\
\texttt{dist\_moe\_router} & new & top-1 mixture-of-experts routing with expert capacity: softmax, argmax, capacity count & tokens & 1 & 2 \\
\bottomrule
\end{tabular}
\end{center}
\caption{The \rev{twenty} bf16 models graded in MLScale, by level: \rev{nine} distributed versions of KernelBench
models~\citep{ouyang2025kernelbench} and eleven new models. \emph{Split} is the dimension the
manifest decomposes over GPUs, and $k$ the degree of the work in it (Appendix~\ref{app:distributed}). Figure~\ref{fig:ml-scaling} labels \texttt{dist\_moe\_router}, \texttt{dist\_sdpa}, \texttt{dist\_all\_to\_all\_transpose} and \texttt{dist\_softmax} as MoE Router, SDPA, All-to-All, and Vocab Softmax.}
\label{tab:exp-ml}
\end{widetable}

\begin{revision}
\textbf{Audits.} Seeded audits in CI repaired the inputs of 201 adopted kernels (all 150 TSVC-2 loops, all 32 PolyBench kernels, and 19 others) and of 24 of our own; on 18 adopted tasks, an answer that ignores its input had passed (on 14 KernelBench problems it still passes KernelBench-Verified's check). CI checks every new kernel for this defect.

\textbf{Optimas and Gymnasium.} Optimas~\citep{wu2025optimas} runs the agent for several rounds within one episode and proposes new instructions from each round's best \texttt{/score} result; its proposer's tokens count toward the limits. A Gymnasium~\citep{towers2024gymnasium} environment over the judge serves optimizers that are not agents: an action is a source with build flags, the observation holds the build log, speedup, and info field, and the reward is the improvement of a correct \texttt{/score} call's speedup over the episode's best.
\end{revision}

\section{Evidence for the capability table}
\label{app:capabilities}

Cells of Table~\ref{tab:capabilities-matrix} come from each benchmark's paper and, where released, its repository.
\emph{Low-level / parallelism reasoning}: tasks built around a low-level optimization or parallelization decision the grade
rewards. \emph{HPC proxy-apps}: multi-kernel code of hundreds to thousands of lines extracted from non-ML production scientific applications, not single operators. \rev{\emph{ML kernels}: a task category of deep-learning operators or models.} \emph{Multi-node}: a timed run can span several nodes. \rev{\emph{Hidden inputs}: the inputs that decide correctness or speed are not shown to the agent in advance, even if they are published. \emph{Held-out final grade}: the final grade runs on inputs not seen during optimization. \emph{Tested comparisons}: a comparison of evaluated systems tested with multiplicity correction. \emph{Interval}: an interval or spread on the evaluated systems' results. \emph{Cost}: the paper reports model cost in tokens or dollars, measured or estimated, for at least one evaluated system or the whole evaluation.}

\section{Experimental setups}
\label{app:experiments}

\subsection{Platform and baselines}
\label{app:setup-details}

\textbf{Platform.} A node holds four MI300A APUs. SGLang~\citep{zheng2024sglang} serves Qwen3.8-27B (FP8) and Kimi-K2.7 with 262{,}144-token contexts; vLLM~0.23.0 (0.27.1 under mini-SWE-agent) serves OSS-120B with 131{,}072.

\textbf{Baselines.} Baselines are Numba for LLR kernels and the fastest of autoparallelized C, sequential C, and parallel Numba for SC kernels; \rev{Pluto (CPU) and PPCG (HIP) are compiler comparators on the LLR track, drawn beside the agents; on the kernels each compiles, agents are $2.4$--$3.0\times$ faster than Pluto (each $p<0.001$) and $2.3$--$5.0\times$ faster than PPCG (significant for Qwen3.8-27B and Kimi-K2.7), each answer timed against the compiler on the same kernel;} the ML track's baseline is PyTorch on one GPU. \rev{The benchmark's hand-tuned Triton~\citep{tillet2019triton} solutions for 80 kernels (40 LLR, 40 SC) are a further GPU reference: over the same Numba build, they reach a geometric-mean speedup of $36.5\times$ on the 39 of 40 LLR kernels the judge accepts (Figure~\ref{fig:packets-scope}, c).}

\begin{widetable}[!htbp]
\centering
\small
\begin{tabular}{@{}p{0.3\textwidth}p{0.62\textwidth}@{}}
\toprule
Intervention & Staged for the agent \\
\midrule
Language skill packet & a skill packet: language, OpenMP, and method pages for the setup's language; a hints page \\
Performance Toolkit & a skill packet: profiling tools; divide-and-conquer, profiling, and compiler optimization-report pages (plus ROCm profiler on AMD) \\
CPF & the kernel's CPF~\citep{budanaz2026cpf} as starting source; a variant instead gives a CPF page and a tool returning the kernel's CPF (Qwen3.8-27B and OSS-120B), reported only in the setups table \\
Git reformulation & the kernel as a git repository with an issue \\
Blind & no score tool, one submission \\
Terse & a page that keeps every agent message brief (Harness20, and C and HIP on LLR-Focus40) \\
Harness & mini-SWE-agent or OpenHands in place of Claude Code, or AutoKernel~\citep{jaber2026autokernel}, an add-on to Claude Code (Harness20) \\
RCCL hints & a skill packet: a page on RCCL collectives (MLScale) \\
\bottomrule
\end{tabular}
\caption{Interventions of \S\ref{sec:interventions}.}
\label{tab:interventions}
\end{widetable}

\textbf{LLR-Focus40.} GPU answers, arrays device-resident, are timed against the CPU Numba build.

\begin{widefigure}[!htbp]
\centering
\includegraphics[width=\widewidth]{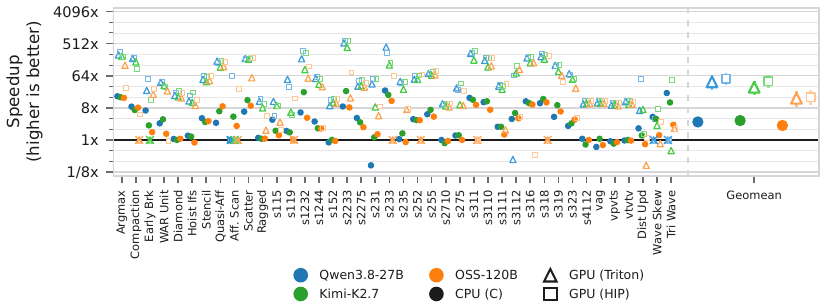}
\caption{Per-kernel speedup on the LLR track, the answers of Figure~\ref{fig:packets-scope}. Geometric means over solved kernels for Qwen3.8-27B, Kimi-K2.7, and OSS-120B: CPU \rev{$3.2$, $3.5$, $2.6\times$; Triton $43.1$, $30.4$, $14.9\times$; HIP $52.6$, $44.5$, $15.7\times$}.}
\label{fig:cheating-per-kernel}
\end{widefigure}

\textbf{\rev{Blind vs.\ Open}.} The same kernels with one submission and no score tool. Figure~\ref{fig:scope} (\rev{c}) pairs \rev{blind and open-submission LLR-Focus40 runs in C for Kimi-K2.7 and OSS-120B}.

\textbf{Git-SciComp.} A generated repository holds one naive starting file, a Makefile, and a performance
issue; the committed starting file is graded. Kernels whose NumPy reference calls BLAS are excluded, as against their
C baseline a speedup would measure library linking.

\textbf{Harness20.} Twenty kernels (Table~\ref{tab:exp-harness20}) in C on the CPU, one episode per kernel. Claude Code is the control; the
treatments are mini-SWE-agent, OpenHands, and AutoKernel, an add-on to Claude Code, on Qwen3.8-27B and OSS-120B.

\textbf{MLScale.} \rev{Twenty} bf16 ML models (Table~\ref{tab:exp-ml}) in single-submission mode, graded under strong and weak scaling
(Appendix~\ref{app:distributed}). Figure~\ref{fig:ml-scaling} shows four of them (MoE Router, SDPA, All-to-All, and Vocab Softmax) \rev{for every setup, with its 42 answers graded on $1$ to $16$ GPUs; one of them, Qwen3.8-27B's All-to-All without skill packets, fails at $16$ GPUs, where PyTorch distributed runs. The RCCL-hint comparison is its own Benjamini--Hochberg family.}

\textbf{Budgets.}\label{app:budgets} Table~\ref{tab:experiments} gives the \rev{limits} per episode. \rev{An episode is never relaunched with a larger budget, and every counted answer stays within these limits. A kernel counts its latest launch with a valid answer, and only that launch's tokens; a cost ratio leaves out kernels without token records. Every answer was re-timed under the final timing protocol (4 inputs, 5 runs each), except a few answers, which keep an earlier grade of the judge, also on hidden inputs.}

\begin{widetable}[!ht]
\begin{center}
\scriptsize
\setlength{\tabcolsep}{3pt}
\begin{tabular}{@{}lll>{\raggedright\arraybackslash}p{1.35cm}l>{\raggedright\arraybackslash}p{1.6cm}>{\raggedright\arraybackslash}p{2.6cm}>{\raggedright\arraybackslash}p{2.3cm}@{}}
\toprule
\textbf{Experiment} & \textbf{Dev.} & \textbf{Harness} & \textbf{Languages} & \textbf{Models} & \textbf{Control} & \textbf{Treatment} & \textbf{Budget} \\
\midrule
LLR-Focus40 & CPU & CC & C\rev{, Fortran} & Q, K, O & no skill packet & Language skill packet & \rev{50M (12M output)} \\
LLR-Focus40 & GPU & CC & HIP, Triton & Q, K, O & no skill packet & Language skill packet & \rev{50M (12M output)} \\
LLR-Focus40 & CPU & CC & C & Q, O & no skill packet & CPF page and tool (variant) & 12M \\
LLR-Focus40 & CPU & CC & C & Q, K, O & no skill packet & CPF as starting source & 12M \\
\rev{Blind vs.\ Open} & CPU & CC & C & \rev{K, O} & open-submission & blind & \rev{50M (12M output)} \\
\rev{LLR blind} & \rev{both} & \rev{CC} & \rev{C, Fortran, HIP} & \rev{Q, K, O} & \rev{no skill packet} & \rev{Language skill packet (HIP: Q, O)} & \rev{50M (12M output)} \\
\rev{SciComp40} & \rev{both} & CC & C\rev{; Fortran, HIP (Q, O)} & Q, K, O & no skill packet & Performance Toolkit & \rev{50M (12M output)} \\
Git-SciComp & CPU & CC & C & Q, K, O & bare kernel & git reformulation & \rev{50M (12M output)} \\
Harness20 & CPU & see text & C & Q, O & CC & mini-SWE-agent, OpenHands, AutoKernel & \rev{48M (12M output)} \\
Terse & both & CC & C, HIP & Q, O & no instruction & brief-message instruction (Harness20: Q, C; LLR-Focus40: C, HIP) & \rev{50M (12M output)} \\
MLScale & GPU & CC & HIP, RCCL & Q, K, O & no skill packet & RCCL hints (Q, O only) & \rev{50M (12M output)} \\
\rev{Blind vs.\ Single} & \rev{both} & \rev{CC} & \rev{C; Fortran (K); HIP (Q, O)} & \rev{Q, K, O} & \rev{single submission} & \rev{blind} & \rev{50M (12M output)} \\
\bottomrule
\end{tabular}
\end{center}
\caption{Setups. Both setups of a pair share model, language, device, and kernels and differ only in
\emph{Treatment}. Q: Qwen3.8-27B, K: Kimi-K2.7, O: OSS-120B.
CC: Claude Code. \emph{Budget}: tokens per episode \rev{(wall time is not compared, Table~\ref{tab:experiments})}.}
\label{tab:exp-setups}
\end{widetable}

\begin{revision}
\subsection{Repeated runs}
\label{app:repeat}

Figure~\ref{fig:temperature}, Table~\ref{tab:repeat-groups}, and Table~\ref{tab:kmp-strategies} show the A/A study of \S\ref{sec:aa}. \rev{Two independent episodes differ by $\sqrt{2}\,\sigma$ in log speedup, where $\sigma$ is the quadratic mean of a model's standard deviations (log speedup: OSS-120B $0.41$, Qwen3.8-27B $1.09$; log tokens: $0.52$ and $0.52$). A paired test on $n$ kernels at $\alpha=0.05$ and 80\% power therefore detects a ratio of $\exp(2.8\sqrt{2}\,\sigma/\sqrt{n})$, with $2.8=z_{0.975}+z_{0.8}$: $1.3\times$ and $2.0\times$ in speed and $1.4\times$ in cost at $n=40$.}

\begin{widetable}[!htbp]\revstart
\centering\scriptsize
\setlength{\tabcolsep}{3pt}
\begin{tabular}{@{}llrrrrrrr@{}}
\toprule
 & & & \multicolumn{4}{c}{Speedup ($\times$)} & \multicolumn{2}{c}{Tokens} \\
\cmidrule(lr){4-7}\cmidrule(l){8-9}
Model & Kernel & $T$ & min & median & max & SD$_{\ln}$ & median (M) & SD$_{\ln}$ \\
\midrule
Qwen3.8-27B & \texttt{kmp} & 1 & 0.51 & 9.65 & 99.5 & 1.73 & 0.92 & 0.63 \\
Qwen3.8-27B & \texttt{kmp} & 0 & 1.83 & 4.19 & 10.9 & 0.57 & 0.45 & 0.43 \\
Qwen3.8-27B & \texttt{addusxx\_g} & 1 & 0.94 & 5.82 & 15.4 & 0.81 & 0.97 & 0.51 \\
Qwen3.8-27B & \texttt{addusxx\_g} & 0 & 2.69 & 6.08 & 10.7 & 0.41 & 0.68 & 0.41 \\
Qwen3.8-27B & \texttt{scan\_affine\_decay} & 1 & 0.04 & 0.99 & 2.00 & 1.03 & 0.70 & 0.53 \\
Qwen3.8-27B & \texttt{scan\_affine\_decay} & 0 & 0.44 & 0.82 & 4.22 & 0.64 & 0.29 & 0.39 \\
OSS-120B & \texttt{kmp} & 1 & 0.85 & 2.22 & 11.2 & 0.67 & 0.15 & 0.71 \\
OSS-120B & \texttt{kmp} & 0 & 1.00 & 2.57 & 70.3 & 1.38 & 0.20 & 0.67 \\
OSS-120B & \texttt{addusxx\_g} & 1 & 0.98 & 2.77 & 3.22 & 0.45 & 0.11 & 0.49 \\
OSS-120B & \texttt{addusxx\_g} & 0 & 1.02 & 2.78 & 3.06 & 0.23 & 0.20 & 0.58 \\
OSS-120B & \texttt{scan\_affine\_decay} & 1 & 0.99 & 0.99 & 0.99 & 0.00 & 0.17 & 0.41 \\
OSS-120B & \texttt{scan\_affine\_decay} & 0 & 0.99 & 0.99 & 1.00 & 0.00 & 0.17 & 0.61 \\
\rev{Qwen3.8-27B} & \rev{\texttt{gelu}} & \rev{1} & \rev{1.07} & \rev{1.33} & \rev{1.45} & \rev{0.08} & \rev{--} & \rev{0.38} \\
\rev{Qwen3.8-27B} & \rev{\texttt{gelu}} & \rev{0} & \rev{0.78} & \rev{1.14} & \rev{1.26} & \rev{0.16} & \rev{--} & \rev{0.48} \\
\rev{OSS-120B} & \rev{\texttt{gelu}} & \rev{1} & \rev{0.71} & \rev{0.76} & \rev{0.98} & \rev{0.10} & \rev{--} & \rev{0.41} \\
\rev{OSS-120B} & \rev{\texttt{gelu}} & \rev{0} & \rev{0.60} & \rev{0.74} & \rev{0.96} & \rev{0.12} & \rev{--} & \rev{0.80} \\
\bottomrule
\end{tabular}
\caption{A/A study: 20 episodes per group; SD$_{\ln}$: standard deviation of the natural log. At the default temperature $T=1$, SD$_{\ln}$ of speedup ranges from near $0$ to $1.7$ ($0.45$--$1.03$ in four of \rev{eight} groups) and of tokens \rev{$0.38$}--$0.71$. \rev{\texttt{gelu} (ML, HIP): speedup over solved episodes (15--19 of 20); token medians omitted, as relaunched episodes restart their count.}}
\label{tab:repeat-groups}
\revend\end{widetable}


\begin{table}[!htbp]\revstart
\centering
\footnotesize
\setlength{\tabcolsep}{3.5pt}
\begin{tabular}{@{}lrrrr@{}}
\toprule
 & \multicolumn{2}{c}{Qwen3.8-27B} & \multicolumn{2}{c}{OSS-120B} \\
\cmidrule(lr){2-3}\cmidrule(l){4-5}
Implemented optimization & Runs & Median & Runs & Median \\
\midrule
Textbook KMP                  & 5 & $0.90\times$  & 6  & $1.00\times$  \\
Brute-force compare + threads & 0 & --            & 1  & \rev{$11.2\times$}  \\
Branch-free automaton         & 3 & $4.17\times$  & 13 & $2.42\times$  \\
Multi-symbol lookup           & 4 & $10.6\times$  & 0  & --            \\
Multi-symbol lookup + threads & 4 & $79.4\times$  & 0  & --            \\
Bit-packed SIMD               & 3 & $11.1\times$  & 0  & --            \\
Bit-packed SIMD + threads     & 1 & $97.8\times$  & 0  & --            \\
\bottomrule
\end{tabular}
\caption{\texttt{kmp} at temperature 1: optimization of each final submission and its median speedup.}
\label{tab:kmp-strategies}
\revend\end{table}
\end{revision}

\begin{revision}
\subsection{Models across tracks and devices}
\label{app:model-gaps}

Two models are compared like two setups: the paired speed ratio over the kernels both solve, with the sign-flip test ($p$ uncorrected). The gaps between models depend on track and device. On LLR in C, Qwen3.8-27B is faster than OSS-120B ($1.26\times$, $p=0.035$), and Qwen3.8-27B against Kimi-K2.7 is inconclusive ($0.92\times$). Qwen3.8-27B's lead over OSS-120B is $2.58\times$ larger in HIP than in C ($1.51$--$4.43$, $p<0.001$, tested on the per-kernel differences between the two devices) and $7.3\times$ on one GPU on the ML track (9 of 9 models, $p=0.004$); on the ML track, Qwen3.8-27B and Kimi-K2.7 are level ($0.97\times$, $p=0.78$). Models agree most on which kernels are hard on LLR (Spearman correlation of per-kernel speedups $0.77$--$0.96$ in C and Fortran), less on SC ($0.58$--$0.71$), and least on the ML track ($0.10$--$0.71$ on one GPU).
\end{revision}

\subsection{\texorpdfstring{\rev{Case study: an agent that reads the input generator}}{Case study: an agent that reads the input generator}}
\label{app:case}

\ifieee\begin{figure}[!ht]\else\begin{widefigure}[H]\fi
\centering\footnotesize
\begin{minipage}[t]{\ifieee\linewidth\else0.24\textwidth\fi}
\begin{lstlisting}[style=codekernel,numbers=none,xleftmargin=0pt,columns=fullflexible,breaklines=false,basicstyle=\figcodefont]
for i in [0:N]:
    if a[i] > K:
        return i, a[i]
return -1, -1
\end{lstlisting}
\codefoot{(a) Baseline}
\end{minipage}\ifieee\par\vspace{4pt}\else\hfill\fi
\begin{minipage}[t]{\ifieee\linewidth\else0.33\textwidth\fi}
\begin{lstlisting}[style=codekernel,numbers=none,xleftmargin=0pt,columns=fullflexible,breaklines=false,basicstyle=\figcodefont]
@reduce(min, first)
doall i in [0:N]:
    if a[i] > K:
        first = min(first, i)
\end{lstlisting}
\codefoot{(b) Parallel}
\end{minipage}\ifieee\par\vspace{4pt}\else\hfill\fi
\begin{minipage}[t]{\ifieee\linewidth\else0.40\textwidth\fi}
\begin{lstlisting}[style=codekernel,numbers=none,xleftmargin=0pt,columns=fullflexible,breaklines=false,basicstyle=\figcodefont]
# first a[i] > K, window first
i = first_above(a, K, 0.4*N, 0.7*N)
if i < 0:      # none: whole array
    i = first_above(a, K, 0, N)
\end{lstlisting}
\codefoot{(c) Window first}
\end{minipage}
\caption{\texttt{ext\_break\_capture}: (a)~the baseline, (b)~a parallel answer, and (c)~\rev{the window-first answer, which is correct only when no match precedes the window}. \texttt{@reduce}: the loop is a reduction.}
\label{fig:case-listings}
\ifieee\end{figure}\else\end{widefigure}\fi

\texttt{ext\_break\_capture} returns the first $a_i>K$ and its index (Figure~\ref{fig:case-listings}a). \rev{A parallel answer searches chunks concurrently and keeps the smallest index found (b). The input generator, which the agent could read, planted the only match in $[0.4N,0.7N)$, a placement the task does not require. An agent that reads this generator can search the window first and fall back to the whole array (c); such an answer is correct on every input the generator produces and wrong on any input with a match before the window. Across setups, 12 final answers searched the window first: 7 of 12 by Qwen3.8-27B, 5 of 9 by Kimi-K2.7, and none of 15 by OSS-120B. On a generator that plants a second, earlier match in every draw, the judge rejects 35 of the 84 graded answers to this kernel, 33 of them on that match (Appendix~\ref{app:protocol}). In the single-submission runs of Blind vs.\ Single, which use that generator, all 7 answers to this kernel are correct ($2.6$--$56\times$).}

\subsection{Second platform: GH200}
\label{app:gh200-transfer}

We regrade the LLR answers on a GH200 cluster: one GH200 module per grade ($72$ Grace cores, one H100), $72$ threads per grade, and the parallel Numba baseline re-timed on the same module. HIP builds through HIP's CUDA backend (ROCm~7.2.4 headers with \texttt{\_\_HIP\_PLATFORM\_NVIDIA\_\_}). An answer is skipped as not portable, never graded as wrong, when it fails an aarch64 syntax probe with an x86-only construct (intrinsics headers, \texttt{\_mm*}, \texttt{\_\_builtin\_ia32\_*}, target attributes), builds only with AMD-only HIP features (rocPRIM or hipCUB, mask-less shuffles, AMD inline assembly, more than 48\,KB static shared memory), or uses AMD-specific Triton options, cache hints, or runtime calls. Table~\ref{tab:gh200} gives the share skipped and the results per language. \rev{Pooled, 583 of 592 CPU and 493 of 512 GPU answers without device-specific code stay correct, with Spearman correlations of $0.92$ and $0.89$.}

\section{Grading and timing}
\label{app:protocol}

\subsection{Secrecy, timing, and evasion}

\textbf{Secrecy.} Two secret seeds fix the sampling: the secret score seed draws the inputs the score tool uses, and the secret submission seed draws the inputs of the final grade; and an image check asserts that no agent image holds either seed. Agents run unprivileged in private user, mount, and PID namespaces and see only their workspace, the kernel and read-only prompt and skill files. \rev{An optional \texttt{/score} info field (build commands, compiler output, gate, timing spread) is off by default; it runs no extra kernel and derives nothing from hidden inputs.}

\textbf{Timed inputs.} Timed repetitions cycle through $D$ seeded input draws \rev{($D=6$; the experiments used $D=4$)}, in one order for both sides; the warmup takes the first draw, so a result cached from the previous call fails the re-check. \rev{\texttt{/score} is timed on two inputs disjoint from \texttt{/submit}'s, three runs each.}

\textbf{Measurement.} One thread per physical core and one GPU per grading process; the judge's clock stops after it synchronizes the timed code's device and OpenMP runtimes.

\textbf{Crediting.} A one-sided Mann--Whitney test~\citep{hoefler2015scientific} at $\alpha=0.1$ in the direction of the medians credits confirmed speedups and confirmed slowdowns alike, and an input it does not confirm is credited $1\times$; it sets $215$ of $8{,}764$ correct timed inputs to $1\times$. \rev{In the experiments, the score tool timed a candidate on one input and reported the fastest of five runs.}

\textbf{Plausibility.}\label{app:evasion}\label{app:evasion-cases} An input is suspect, and left out of $S_i$, if its median speedup exceeds $2000\times$ (host) or $16{,}000\times$ (device), its time is below declared bytes over twice the $5.3$\,TB/s HBM peak, or a device check fires; an answer whose inputs are all suspect is credited $1\times$. Section~\ref{sec:evasion} describes the evasions these checks caught.

\subsection{Correctness}

\begin{widetable}[!ht]
\begin{center}
\small
\begin{tabular}{lcccccc}
\toprule
 & fp64 & fp32 & fp16 & bf16 & fp8 (e4m3) & fp8 (e5m2) \\
\midrule
\texttt{rtol} & $10^{-9}$  & $10^{-3}$ & $10^{-2}$ & $3\cdot10^{-2}$ & \rev{$0.125$} & \rev{$0.25$} \\
\texttt{atol} & $10^{-11}$ & $10^{-5}$ & $10^{-3}$ & $10^{-2}$       & $0.125$   & $0.25$ \\
\bottomrule
\end{tabular}
\end{center}
\caption{Validation bands by precision. Other formats: $\texttt{rtol}=\sqrt{\varepsilon}$ clamped to $[10^{-11},0.25]$,
$\texttt{atol}=\max(10^{-2}\,\texttt{rtol},\varepsilon)$. \rev{fp8 accumulates in fp32~\citep{blanchard2020mixed}, bf16 and fp16 in their own precision. The experiments accumulated bf16 and fp16 in fp32 and used an fp8 \texttt{rtol} of $10^{-1}$ and $2\cdot10^{-1}$.}}
\label{tab:tolerances}
\end{widetable}

\textbf{Bands.} NaN and $\pm\infty$ positions must agree, and each output's \texttt{atol}
(Table~\ref{tab:tolerances}) is raised to the reassociation floor
\begin{equation}
\texttt{atol}_{\text{eff}} \;=\; \max\!\big(\texttt{atol},\;\varepsilon_{\text{acc}}\sqrt{\ell}\,\lVert x_{\text{ref}}\rVert_\infty\big),
\end{equation}
where $\varepsilon_{\text{acc}}$ is the accumulation precision's machine epsilon and $\ell$ the largest product of
an input's size symbols absent from the written output shape ($K$ for $(M,K)\times(K,N)$). Kernels whose shapes do not fix
$\ell$ use their largest input's element count; sequential scans declare their chain length. A configuration with $\varepsilon_{\text{acc}}\sqrt{\ell}\ge\texttt{rtol}$ is
ungradeable, not incorrect\rev{; this refusal applies only to fp64, fp32, and fp8; bf16 and fp16 use the bound of a tree reduction, $\tfrac{1}{2}\varepsilon_{\text{acc}}(\lceil\log_2\ell\rceil+1)\lVert x_{\text{ref}}\rVert_\infty$, in place of the $\sqrt{\ell}$ term}.

\textbf{Determinism.} Two runs of one binary agree when their largest elementwise difference over
$\varepsilon\sqrt{\ell}$ times the first run's max-norm is below $30$, LAPACK's test-ratio
threshold~\citep{blackford1999lawn41}.

\begin{revision}
\textbf{All checks.} Table~\ref{tab:checks} lists the checks of a grade.

\begin{widetable}[!htbp]\revstart
\centering\footnotesize
\setlength{\tabcolsep}{4pt}
\begin{tabular}{@{}llp{0.51\textwidth}@{}}
\toprule
Check & Applies to & Passes when \\
\midrule
Exact match & integer and boolean outputs & $x=x_{\text{ref}}$ elementwise \\
Special values & floating-point outputs & NaN and $\pm\infty$ positions agree \\
Elementwise band & floating-point outputs & $|x-x_{\text{ref}}|\le\texttt{atol}_{\text{eff}}+\texttt{rtol}\,|x_{\text{ref}}|$, $\texttt{atol}_{\text{eff}}=\max(\texttt{atol},\,\varepsilon_{\text{acc}}\sqrt{\ell}\,\lVert x_{\text{ref}}\rVert_\infty)$ \\
Whole output & floating-point outputs, all ranks & $\lVert x-x_{\text{ref}}\rVert_2\le\tau\,\lVert x_{\text{ref}}\rVert_2$ \\
Gradeable & each held-out case & rounding alone stays below the band: $\varepsilon_{\text{acc}}\sqrt{\ell}<\texttt{rtol}$ at that case's sizes \\
Determinism & two runs on the public input & the runs agree, $\lVert x_1-x_2\rVert_\infty\le30\,\varepsilon_{\text{acc}}\sqrt{\ell}\,\lVert x_1\rVert_\infty$, and each passes the checks above \\
Credit & each timed input & one-sided Mann--Whitney $p<0.1$ \\
Plausible speed & each timed input & median speedup $\le2000\times$ (host), $\le16{,}000\times$ (device) \\
Plausible time & each timed input & time $\ge$ declared bytes$/(2\cdot5.3$\,TB/s$)$, twice the MI300A's HBM peak \\
\bottomrule
\end{tabular}
\caption{\rev{Checks a release grade applies to an answer. $\ell$: number of terms summed into one output element (1 for elementwise kernels); $\varepsilon_{\text{acc}}$: machine epsilon of the accumulation precision; $\tau=\texttt{rtol}$; 30: LAPACK's test-ratio threshold.}}
\label{tab:checks}
\revend\end{widetable}

\textbf{Tolerance study.} Over 17 corruptions of every kernel's double-precision output, the experiments' tolerance accepts a wrong output in at most $2.1\%$ (LLR) and $7.0\%$ (SC) of cases; with the whole-output check, the SC rate falls to at most $3.2\%$, while correct \texttt{-O3} builds reach at most $0.038$ of the check's bound.
\end{revision}

\subsection{Scoring rules}

\begin{revision}
\textbf{Efficacy statistics.}\label{app:efficacy-statistics} Table~\ref{tab:protocol-defaults} gives the protocol's defaults\rev{; every comparison in this paper uses its tests, and} margins and $\alpha$ are configurable.

\begin{table}[!htbp]\revstart
\centering\footnotesize
\setlength{\tabcolsep}{4pt}
\begin{tabular}{@{}llp{\ifieee0.5\linewidth\else0.66\linewidth\fi}@{}}
\toprule
Part & Setting & Default \\
\midrule
Design & Kernels & the same kernels for both setups, drawn at random within a fixed quota per level \\
 & Episodes & one scored episode per kernel and setup \\
 & Limits & one token and time limit for both setups \\
 & Inputs & hidden from the agent; final grade on held-out inputs \\
\midrule
Inference & Speed, cost & paired sign-flip permutation test on the mean log ratio $\bar d$, $\rho=e^{\bar d}$: all $2^n$ sign vectors up to $n=16$, else 19{,}999 seeded draws; kernels with $d_i=0$ stay in \\
 & Solve rate & exact McNemar test on $g$ and $l$; pooled: Tango's score test and interval~\citep{tango1998equivalence} \\
 & Correction & Benjamini--Hochberg procedure, false discovery rate 5\% per experiment and measure; difference and equivalence corrected separately \\
 & Equivalence & two one-sided tests (TOST) on the 90\% confidence interval, margins $1.2\times$ (ratios) and 10 points (solve rate); a significant difference takes precedence \\
\midrule
Output & Per comparison & each ratio with its 95\% confidence interval; verdict: different, equivalent, or inconclusive; minimum detectable effect at 80\% power; token counts per component \\
\bottomrule
\end{tabular}
\caption{Default efficacy protocol.}
\label{tab:protocol-defaults}
\revend\end{table}

\textbf{All comparisons.}\label{app:all-comparisons} Of the \rev{48} comparisons (Table~\ref{tab:all-comparisons}), \rev{2} compare blind against open submission; \rev{6} show a significant speed effect\rev{, 5 an equivalent speed,} and \rev{9} a significant cost effect.

\begin{widetable}[!p]\revstart
\centering\tiny
\setlength{\tabcolsep}{2.5pt}
\begin{tabular}{@{}llllrrrlllrr@{}}
\toprule
Intervention & Experiment & Model & Lang. & $n$ & Solved & $\Delta R$ & $\rho_S$ [95\%] & $\rho_C$ [95\%] & $R$/$S$/$C$ & MDE$_S$ & MDE$_C$ \\
\midrule
Language packet & LLR CPU & Qwen3.8-27B & C & 40 & 39/39 & 0 & 1.22 [1.05, 1.43] & 1.02 [0.78, 1.33] & i/d/i & 1.25 & 1.47 \\
Language packet & LLR CPU & Qwen3.8-27B & Fortran & 40 & 39/40 & +2.5 & 1.11 [0.92, 1.34] & 1.39 [1.10, 1.77] & i/i/d & 1.31 & 1.40 \\
Language packet & LLR CPU & OSS-120B & C & 40 & 40/39 & -2.5 & 1.03 [0.95, 1.11] & 0.90 [0.68, 1.19] & i/e/i & 1.12 & 1.50 \\
Language packet & LLR CPU & OSS-120B & Fortran & 40 & 39/38 & -2.5 & 0.93 [0.80, 1.07] & 0.74 [0.55, 0.99] & i/i/i & 1.24 & 1.52 \\
Language packet & LLR CPU & Kimi-K2.7 & C & 40 & 40/39 & -2.5 & 1.00 [0.89, 1.13] & 1.21 [1.02, 1.45] & i/e/i & 1.19 & 1.29 \\
Language packet & LLR CPU & Kimi-K2.7 & Fortran & 40 & 39/38 & -2.5 & 1.21 [1.09, 1.35] & 1.14 [0.96, 1.35] & i/d/i & 1.17 & 1.27 \\
CPF & LLR CPU & Qwen3.8-27B & C & 40 & 40/37 & -7.5 & 1.11 [0.90, 1.37] & 2.05 [1.45, 2.92] & i/i/d & 1.41 & 1.59 \\
CPF & LLR CPU & OSS-120B & C & 40 & 38/38 & 0 & 1.23 [1.08, 1.39] & 0.97 [0.68, 1.41] & i/d/i & 1.26 & 1.62 \\
CPF & LLR CPU & Kimi-K2.7 & C & 40 & 40/39 & -2.5 & 0.92 [0.73, 1.15] & 0.98 [0.74, 1.29] & i/i/i & 1.42 & 1.48 \\
Language packet & LLR GPU & Qwen3.8-27B & HIP & 40 & 39/37 & -5.0 & 0.96 [0.89, 1.02] & 0.95 [0.77, 1.16] & i/e/i & 1.10 & 1.34 \\
Language packet & LLR GPU & Qwen3.8-27B & Triton & 40 & 37/39 & +5.0 & 1.10 [0.85, 1.49] & 1.08 [0.95, 1.23] & i/i/i & 1.62 & 1.20 \\
Language packet & LLR GPU & OSS-120B & HIP & 40 & 39/39 & 0 & 1.28 [0.89, 1.84] & 0.93 [0.72, 1.21] & i/i/i & 1.68 & 1.44 \\
Language packet & LLR GPU & OSS-120B & Triton & 40 & 33/32 & -2.5 & 0.79 [0.50, 1.19] & 0.84 [0.61, 1.17] & i/i/i & 1.97 & 1.58 \\
Language packet & LLR GPU & Kimi-K2.7 & HIP & 40 & 39/40 & +2.5 & 1.22 [0.98, 1.53] & 1.11 [0.90, 1.38] & i/i/i & 1.39 & 1.36 \\
Language packet & LLR GPU & Kimi-K2.7 & Triton & 40 & 38/39 & +2.5 & 1.14 [0.91, 1.44] & 0.92 [0.75, 1.14] & i/i/i & 1.41 & 1.35 \\
Language packet & LLR blind & Qwen3.8-27B & C & 40 & 36/37 & +2.5 & 1.01 [0.84, 1.22] & 1.04 [0.79, 1.38] & i/i/i & 1.31 & 1.48 \\
Language packet & LLR blind & Qwen3.8-27B & Fortran & 40 & 40/39 & -2.5 & 1.09 [0.92, 1.29] & 1.10 [0.82, 1.48] & i/i/i & 1.28 & 1.53 \\
Language packet & LLR blind & Qwen3.8-27B & HIP & 40 & 37/40 & +7.5 & 0.89 [0.53, 1.53] & 0.95 [0.80, 1.13] & i/i/i & 2.19 & 1.28 \\
Language packet & LLR blind & OSS-120B & C & 40 & 38/37 & -2.5 & 1.06 [0.94, 1.20] & 0.92 [0.73, 1.16] & i/i/i & 1.19 & 1.39 \\
Language packet & LLR blind & OSS-120B & Fortran & 40 & 35/35 & 0 & 0.87 [0.69, 1.09] & 0.79 [0.65, 0.97] & i/i/i & 1.39 & 1.33 \\
Language packet & LLR blind & OSS-120B & HIP & 40 & 34/32 & -5.0 & 0.92 [0.66, 1.31] & 0.78 [0.61, 0.99] & i/i/i & 1.66 & 1.42 \\
Language packet & LLR blind & Kimi-K2.7 & C & 40 & 37/37 & 0 & 1.05 [0.88, 1.27] & 1.01 [0.82, 1.24] & i/i/i & 1.30 & 1.34 \\
Language packet & LLR blind & Kimi-K2.7 & Fortran & 40 & 38/39 & +2.5 & 1.06 [0.91, 1.23] & 1.19 [0.91, 1.55] & i/i/i & 1.25 & 1.46 \\
Blind & LLR & Kimi-K2.7 & C & 40 & 40/37 & -7.5 & 0.83 [0.73, 0.95] & 2.65 [2.04, 3.45] & i/d/d & 1.21 & 1.45 \\
Blind & LLR & OSS-120B & C & 40 & 40/38 & -5.0 & 0.94 [0.82, 1.07] & 1.08 [0.78, 1.50] & i/i/i & 1.20 & 1.60 \\
Git framing & Git-SciComp & Qwen3.8-27B & C & 27 & 27/26 & -3.7 & 1.15 [0.73, 1.77] & 0.87 [0.67, 1.14] & i/i/i & 1.88 & 1.47 \\
Git framing & Git-SciComp & OSS-120B & C & 27 & 21/26 & +18.5 & 0.90 [0.73, 1.10] & 0.62 [0.40, 0.95] & i/i/d & 1.34 & 1.86 \\
Git framing & Git-SciComp & Kimi-K2.7 & C & 27 & 26/26 & 0 & 0.78 [0.70, 0.88] & 1.99 [1.58, 2.52] & i/d/d & 1.18 & 1.39 \\
Perf.\ toolkit & SciComp40 & Qwen3.8-27B & C & 40 & 39/35 & -10.0 & 1.19 [0.80, 1.73] & 1.10 [0.87, 1.39] & i/i/i & 1.78 & 1.39 \\
Perf.\ toolkit & SciComp40 & OSS-120B & C & 40 & 30/37 & +17.5 & 1.00 [0.91, 1.10] & 0.98 [0.70, 1.38] & i/e/i & 1.15 & 1.63 \\
Perf.\ toolkit & SciComp40 & Kimi-K2.7 & C & 40 & 38/36 & -5.0 & 0.72 [0.54, 0.97] & 2.29 [1.82, 2.87] & i/i/d & 1.51 & 1.38 \\
Perf.\ toolkit & SciComp40 & Qwen3.8-27B & Fortran & 40 & 32/34 & +5.0 & 1.57 [1.09, 2.28] & 1.22 [0.99, 1.49] & i/i/i & 1.70 & 1.34 \\
Perf.\ toolkit & SciComp40 & OSS-120B & Fortran & 40 & 30/28 & -5.0 & 0.95 [0.79, 1.15] & 1.28 [0.87, 1.86] & i/i/i & 1.31 & 1.71 \\
Perf.\ toolkit & SciComp40 & Qwen3.8-27B & HIP & 40 & 31/30 & -2.5 & 0.64 [0.41, 0.92] & 0.96 [0.82, 1.11] & i/d/i & 1.94 & 1.24 \\
Perf.\ toolkit & SciComp40 & OSS-120B & HIP & 40 & 29/26 & -7.5 & 0.86 [0.58, 1.26] & 1.04 [0.68, 1.61] & i/i/i & 1.74 & 1.84 \\
mini-SWE-agent & Harness20 & Qwen3.8-27B & C & 20 & 20/19 & -5.0 & 1.48 [0.92, 2.39] & 0.96 [0.70, 1.31] & i/i/i & 1.95 & 1.56 \\
OpenHands & Harness20 & Qwen3.8-27B & C & 20 & 20/19 & -5.0 & 1.01 [0.72, 1.41] & 1.59 [0.98, 2.60] & i/i/i & 1.60 & 1.99 \\
AutoKernel & Harness20 & Qwen3.8-27B & C & 20 & 20/20 & 0 & 1.24 [0.79, 1.95] & 1.06 [0.80, 1.40] & i/i/i & 1.91 & 1.49 \\
mini-SWE-agent & Harness20 & OSS-120B & C & 20 & 19/17 & -10.0 & 0.99 [0.65, 1.53] & 1.35 [0.98, 1.87] & i/i/i & 1.85 & 1.60 \\
OpenHands & Harness20 & OSS-120B & C & 20 & 19/16 & -15.0 & 1.09 [0.78, 1.56] & 0.45 [0.29, 0.67] & i/i/d & 1.65 & 1.80 \\
AutoKernel & Harness20 & OSS-120B & C & 20 & 19/20 & +5.0 & 0.82 [0.57, 1.16] & 0.87 [0.59, 1.29] & i/i/i & 1.66 & 1.75 \\
Terse & LLR & Qwen3.8-27B & C & 40 & 39/40 & +2.5 & 1.06 [0.89, 1.27] & 0.75 [0.59, 0.95] & i/i/d & 1.29 & 1.40 \\
Terse & LLR & Qwen3.8-27B & HIP & 40 & 39/40 & +2.5 & 1.03 [0.93, 1.13] & 0.94 [0.77, 1.15] & i/e/i & 1.15 & 1.33 \\
Terse & LLR & OSS-120B & C & 40 & 40/40 & 0 & 0.93 [0.82, 1.05] & 1.54 [1.14, 2.08] & i/i/d & 1.19 & 1.53 \\
Terse & LLR & OSS-120B & HIP & 40 & 39/39 & 0 & 1.06 [0.67, 1.66] & 1.20 [0.93, 1.54] & i/i/i & 1.91 & 1.43 \\
Terse & Harness20 & Qwen3.8-27B & C & 20 & 20/20 & 0 & 1.21 [0.78, 1.94] & 1.28 [0.96, 1.70] & i/i/i & 1.93 & 1.50 \\
RCCL hints & MLScale & Qwen3.8-27B & HIP & 20 & 15/12 & -15.0 & 0.87 [0.41, 1.55] & 1.05 [0.83, 1.34] & i/i/i & 2.69 & 1.40 \\
RCCL hints & MLScale & OSS-120B & HIP & 20 & 11/10 & -5.0 & 1.02 [0.24, 7.53] & 0.81 [0.53, 1.24] & i/i/i & 9.32 & 1.82 \\
\bottomrule
\end{tabular}
\caption{All \rev{48} comparisons, \rev{2} of them blind against open submission, \rev{every one except CPF from runs under the limits of Table~\ref{tab:experiments},} under the tests of Appendix~\ref{app:efficacy-statistics}. Solved: control/treated; $\Delta R$: solve-rate change in points; verdicts for solve rate, speed, and cost: d different, e equivalent, i inconclusive; MDE$_S$, MDE$_C$: smallest speed and cost ratio detectable at 80\% power. Intervals are uncorrected; verdicts control the \rev{false discovery rate at 5\% (Benjamini--Hochberg) within each experiment and measure}. $\rho_C>1$: treated cheaper. Blind rows compare blind with open submission.}
\label{tab:all-comparisons}
\revend\end{widetable}

\begin{widetable}[!htbp]\revstart
\centering\scriptsize
\setlength{\tabcolsep}{1.6pt}
\begin{tabular}{@{}lrlllllr@{}}
\toprule
Intervention & Comp. & Speed (95\%) & Verdict & Solve rate (95\%) & Verdict & Cost (95\%) & Verdict  \\
\midrule
Language packet & 20 & $1.05\times$ ($0.99$--$1.11$) & equiv. & $-0.1$ ($-2.0$ to $1.8$) & equiv. & $0.99\times$ ($0.94$--$1.04$) & equiv.  \\
Blind vs.\ Open & 2 & $0.88\times$ ($0.81$--$0.97$) & diff. & $-6.2$ ($-13.8$ to $-1.4$) & incon. & $1.69\times$ ($1.35$--$2.13$) & diff.  \\
Blind vs.\ Single, SciComp40 & 3 & $0.81\times$ ($0.69$--$0.94$) & diff. & $-8.3$ ($-17.2$ to $0.3$) & incon. & $1.78\times$ ($1.49$--$2.12$) & diff.  \\
Blind vs.\ Single, Focus40 & 9 & $0.86\times$ ($0.78$--$0.95$) & diff. & $-5.6$ ($-9.2$ to $-2.2$) & diff. & $1.42\times$ ($1.30$--$1.55$) & diff.  \\
Git framing & 3 & $0.94\times$ ($0.79$--$1.11$) & incon. & $+4.9$ ($-2.3$ to $13.2$) & incon. & $1.02\times$ ($0.83$--$1.26$) & incon.  \\
Perf.\ toolkit & 7 & $0.96\times$ ($0.84$--$1.09$) & incon. & $-1.1$ ($-5.9$ to $3.7$) & equiv. & $1.21\times$ ($1.08$--$1.36$) & diff.  \\
CPF$^{\dagger}$ & 3 & $1.07\times$ ($0.96$--$1.20$) & incon. & $-3.3$ ($-8.3$ to $-0.1$) & equiv. & $1.29\times$ ($1.05$--$1.59$) & diff.  \\
Harnesses & 4 & $1.14\times$ ($0.94$--$1.38$) & incon. & $-8.8$ ($-18.0$ to $-0.7$) & incon. & $0.98\times$ ($0.79$--$1.21$) & incon.  \\
AutoKernel plugin & 2 & $1.02\times$ ($0.76$--$1.36$) & incon. & $+2.5$ ($-6.5$ to $12.9$) & incon. & $0.96\times$ ($0.76$--$1.21$) & incon.  \\
Terse & 5 & $1.04\times$ ($0.92$--$1.17$) & equiv. & $+1.1$ ($-1.6$ to $4.3$) & equiv. & $1.09\times$ ($0.97$--$1.23$) & incon.  \\
\bottomrule
\end{tabular}
\caption{Pooled effect of each intervention, kernels as the unit; cost ratio above $1\times$: cheaper; solve rates in points. Verdicts: Benjamini--Hochberg over the \rev{nine rows without CPF, per measure} (Table~\ref{tab:protocol-defaults}). \rev{Every row except CPF comes from runs under the limits of Table~\ref{tab:experiments}; $^{\dagger}$CPF keeps its first runs. Blind vs.\ Open: Kimi-K2.7 $0.83\times$, OSS-120B $0.94\times$.}}
\label{tab:pooled}
\revend\end{widetable}
\end{revision}

\textbf{Example.}\label{app:efficacy-example} Illustrative speedups, $\textsf{X}$ unsolved:
\begin{equation}
\begin{gathered}
S^{\text{before}}=(2,\,4,\,\textsf{X},\,1,\,8,\,2,\,4,\,1),\eqbreak
S^{\text{after}}=(4,\,4,\,2,\,0.5,\,16,\,2,\,8,\,1).
\end{gathered}
\label{eq:efficacy-example}
\end{equation}
Solved kernels rise from $7$ to $8$ ($\rho_R=1.14$; $g=1$, $l=0$); over the seven both solve, the geometric-mean
speedup rises from $2.44$ to $2.97$ ($\rho_S=1.22$\rev{; 95\% sign-flip interval $[0.71,\,2.00]$, exact $p=0.63$}).

\textbf{Token counts.}\label{app:token-cost-rules} Engine cache counters are not used for
\eqref{eq:token-cost}. Without a server count (killed process) the model's tokenizer counts the transcript,
undercounting 2--4\%. After compaction the
rebuilt prompt counts as input.

\section{Distributed execution and scaling}
\label{app:distributed}

\textbf{Weak-scaling sizes.} In \eqref{eq:scaling}, $P$ counts PEs, one MPI rank per PE. A manifest names the decomposed size symbols $N=(N_1,\dots,N_d)$ and the
degree $k$ of the work in them, $W(sN)=s^kW(N)$ (a GEMM partitioning one row axis has $k=1$). Fixed work per PE needs
$s=P^{1/k}$, an integer $m$ at $P=m^k$:
\begin{equation}
N_{P,j} \;=\; m\,N_{1,j},\qquad j=1,\dots,d,\qquad P=m^{k},
\end{equation}
so $r_i(P)=P$, as in isoefficiency~\citep{grama1993isoefficiency}, HPCG and LULESH~\citep{heroux2013hpcg,karlin2013lulesh}.
At other $P$ the symbols scale by $P^{1/k}$, rounded, and $r_i(P)=W(N_P)/W(N_1)$ is recorded. Table~\ref{tab:distributed-motifs} gives the work and the exact weak-scaling PE counts per motif.

\begin{widetable}[!ht]
\centering
\footnotesize
\begin{tabular}{@{}llll@{}}
\toprule
Motif & Work & Weak $P$ & Strong decomposition \\
\midrule
Dense LA / GEMM      & $O(N^{3})$                          & $m^{3}$ & 2D block-cyclic \\
2D stencil           & $O(N^{2}T)$                         & $m^{2}$ & 1D / 2D block \\
3D stencil           & $O(N^{3}T)$                         & $m^{3}$ & 3D block \\
SpMV / BFS / map-reduce & $O(\mathrm{nnz})$ / $O(V{+}E)$ / $O(N)$ & any ($k{=}1$) & row / graph partition \\
Direct $N$-body      & $O(N^{2})$                          & $m^{2}$ & particle / domain \\
FFT                  & $O(N \log N)$                       & strong only$^{\dagger}$ & slab / pencil transpose \\
\bottomrule
\end{tabular}
\caption{Work, exact weak-scaling PE counts, and strong-scaling decomposition per motif. $^{\dagger}$No integer growth multiplies $N\log N$ exactly by $P$.}
\label{tab:distributed-motifs}
\end{widetable}

\textbf{Ownership.} \rev{A submission can request each input's precision and layout; the judge converts the data outside the timed region, grades at the task's own precision, and grades distributed outputs from per-rank sums.} A kernel declares each array's layout: a ScaLAPACK~\citep{blackford1997scalapack} block, cyclic
or block-cyclic distribution over a process grid, or replication. The agent writes all communication; the drivers only scatter
inputs and gather outputs, so the correctness check and timed region keep their single-node definitions.

\textbf{MLScale.} The grade measures on $1$, $2$, $4$, $8$ and $16$ GPUs across four nodes, four GPUs per node, and one rank per GPU; a point is the median over runs of the slowest GPU's time, and $B_i$ is the median of PyTorch (\texttt{torch.compile}) on one GPU at the base size. Input
dimensions and per-GPU blocks are multiples of $64$, the strong-scaling split dimension a multiple of $1024$.

\section{Token-cost conventions of prior work}
\label{app:token-cost}

Where prior work states how it counts tokens, it
sums every request's prompt, so in multi-turn agent sessions cached input dominates the
count~\citep{yu2026evolvingensemble,zhu2026tracelab} and varies with the
serving engine's cache policy~\citep{zheng2024sglang,lumer2026dontbreakcache}.

\section{Reproducing the results}
\label{app:reproduce}
\begingroup\raggedright\Urlmuskip=0mu\relax

The only release is one anonymized Zenodo zip (\artifacturl, 31\,MB) with a top-level \texttt{README.md} and two folders:
\begin{itemize}[leftmargin=*,noitemsep,topsep=2pt]
\item \texttt{reproducibility/}, the artifact repository: the pipeline scripts (\texttt{download.sh}, \texttt{run\_all.sh}, \texttt{stats.sh}, \texttt{fig*.sh}, \texttt{tab3.sh}, \texttt{lib/}, \texttt{tools/}), the measurements, extracted in \texttt{data/} and as \texttt{data.tar.zst} with one checksum per file in \texttt{DATA\_SHA256SUMS}, the pooled answers in \texttt{work/}, every table and figure, the checksums of all outputs in \texttt{SHA256SUMS}, the submitted sources behind \S\ref{sec:evasion} and \S\ref{sec:llr-behaviour} (\texttt{case-studies/}) and every version of the skill packets (\texttt{skill\_histories/});
\item \texttt{hpcagent-bench/}, the benchmark at the release tagged \texttt{paper-experiments}: benchmark code, experiment scripts and container definitions, with DaCe pinned through the \texttt{dace} dependency group of \texttt{pyproject.toml}.
\end{itemize}
User names, paths, host names and commit ids are replaced by placeholders, which changes no number.

\subsection{Regenerating every figure, table and number}
\label{app:reproduce-figures}

This needs no GPU and no model access, only Linux with Python~$\ge3.12$. From the unpacked zip:

\begin{lstlisting}[style=codekernel,numbers=none,xleftmargin=0pt,columns=fullflexible,mathescape=false]
cd reproducibility
python3 -m venv .venv && . .venv/bin/activate
pip install -e ../hpcagent-bench
./run_all.sh
\end{lstlisting}

The measurements ship extracted; \texttt{DATA\_URL=file://\$PWD/data.tar.zst ./download.sh} re-extracts them and fails unless every file matches \texttt{DATA\_SHA256SUMS}. \texttt{run\_all.sh} pools every answer at its final grade, runs the paired tests of \S\ref{sec:score}, redraws every figure and table with the benchmark's plotting library, and checks all $54$ outputs against \texttt{SHA256SUMS}. It ends with \texttt{OK: 54 files match SHA256SUMS}, or \texttt{sha256sum} names each file that differs. Table~\ref{tab:repro-scripts} lists the script behind each output.

\begin{table}[ht]
\centering\ifieee\footnotesize\setlength{\tabcolsep}{3pt}\else\small\fi
\fitcolumn{\begin{tabular}{@{}lll@{}}
\toprule
Script & Output & In the paper \\
\midrule
\texttt{stats.sh} & pooled answers, paired-test tables (\texttt{tables/*.csv}) & every figure \\
\texttt{fig2.sh} & \texttt{figures/efficacy\_packets\_and\_scope.pdf} & Figure~\ref{fig:packets-scope} \\
\texttt{fig3.sh} & \texttt{figures/scope\_row.pdf} & Figure~\ref{fig:scope} \\
\texttt{fig4.sh} & \texttt{figures/cost\_weighting.pdf} & Figure~\ref{fig:cost-weighting} \\
\texttt{fig5.sh} & \texttt{figures/ml\_scaling.pdf} & Figure~\ref{fig:ml-scaling} \\
\texttt{fig8.sh} & \texttt{figures/cheating\_per\_kernel.pdf} & Figure~\ref{fig:cheating-per-kernel} \\
\texttt{tab3.sh} & \texttt{tables/gh200-transfer-table.tex} & Table~\ref{tab:gh200} \\
\bottomrule
\end{tabular}}
\caption{Scripts of \texttt{reproducibility}; each runs after \texttt{stats.sh}.}
\label{tab:repro-scripts}
\end{table}

\subsection{Rerunning an experiment}
\label{app:reproduce-experiments}

\textbf{Platform.} The experiments ran on a Slurm~25.05 cluster of MI300A nodes (four APUs per node, each $24$ Zen~4 cores and $228$ CDNA~3 compute units sharing $128$\,GB HBM3) with the Enroot container runtime; the GH200 regrade ran on GH200 nodes (four superchips per node, each $72$ Neoverse~V2 cores and one H100). The CPU experiments need CPU nodes only; the GPU experiments need AMD GPUs. All toolchains live in the images: ROCm~7.2.0 in the judge, agent and model-server images (SGLang for Qwen3.8-27B and Kimi-K2.7, vLLM for OSS-120B).

\textbf{One-time setup.} Two jobs of \path{hpcagent-bench/containers/cluster/ce-images} prepare the cluster (commands in its \texttt{README.md}): \texttt{build\_and\_verify.sbatch} builds the judge, agent and model-server images and checks each against the device, and \texttt{inference/fetch\_weights.sbatch} downloads the model weights. The CPF setups also read their pre-rendered forms from \texttt{HPCAGENT\_BENCH\_CPF\_PRERENDER\_DIR}.

\textbf{One command per experiment.} Set \texttt{SCRATCH} to the directory that holds \texttt{hpcagent-bench/} and its virtual environment. From \texttt{hpcagent-bench/experiments/}, the command of Table~\ref{tab:rerun} reruns one experiment end to end, with the budgets of Table~\ref{tab:experiments}; \texttt{SUBMIT=0} prints the plan without submitting. It submits one Slurm job per setup, and each job starts the model server, the judge and the agents in the fixed images. Every job ends within its wall limit of at most $23$\,h, so graded results are back within a day when the cluster starts an experiment's jobs together.

\begin{table}[ht]
\centering\ifieee\footnotesize\setlength{\tabcolsep}{3pt}\else\small\fi
\fitcolumn{\begin{tabular}{@{}ll@{}}
\toprule
Experiment & \rev{Commands} (from \texttt{experiments/}) \\
\midrule
\rev{LLR-Focus40} & \rev{\path{studies/rev-llr40.md}} \\
\rev{CPF (first runs)} & \texttt{SUBMIT=1 ./submit-cpf-llr40.sh} \\
\rev{Blind vs.\ Open, Blind vs.\ Single (LLR)} & \rev{\path{studies/llr40-blind.md}} \\
\rev{SciComp40, Blind vs.\ Single (SC)} & \rev{\path{studies/rev-scicomp40.md}} \\
Git-SciComp & \rev{\path{studies/rev-gitscicomp40.md}} \\
Harness20 & \rev{\path{studies/harness20.md}} \\
\rev{MLScale} & \rev{\path{studies/mlscale20.md}} \\
\rev{A/A study (ML)} & \rev{\path{studies/rev-repeat3-ml-feedback.md}} \\
\rev{LLR baselines (no agent)} & \rev{\texttt{SUBMIT=1 ./submit-canon-llr40.sh}} \\
\bottomrule
\end{tabular}}
\caption{\rev{Where each experiment's launch commands are, from \texttt{experiments/}. \texttt{studies/revision.md} gives the one limit of every run and which earlier runs are reused; \texttt{statistics/revision/} holds the scripts behind Tables~\ref{tab:all-comparisons} and~\ref{tab:pooled}.}}
\label{tab:rerun}
\end{table}

\textbf{Grading.} Every submit script chains a grading job on each agent job (\texttt{finalize\_grade.sbatch}), which applies the current grading rules, the ones behind every number in this paper. The final grade re-times each answer on inputs drawn from the secret submission seed (Appendix~\ref{app:protocol}); items a grading job does not reach within its wall time are planned again by the next one (\texttt{regrade\_rest.py}). An answer whose final re-timing yields no valid measurement keeps its last valid grade, and an answer without a final grade keeps the grade the judge gave during the episode. Kernels a setup never ran are excluded from that setup; a kernel that ran and failed counts as unsolved, so it enters the success rate and the token cost but not $\rho_S$. The multi-node scaling grade of the ML track is scheduled as its answers arrive.

\textbf{From finished runs to figures.} Each job writes its run directory to \path{$SCRATCH/hpcagent-bench-runs/<experiment>-<date>/<job>}. On the cluster, from \texttt{reproducibility/}, \texttt{PULL=0 MIRROR=\$SCRATCH ./download.sh --from-cluster} rebuilds \texttt{data/} from those directories; elsewhere, \texttt{./download.sh --from-cluster} first mirrors them over ssh from the host set in \texttt{tools/cluster.env} (copied from \texttt{tools/cluster.env.example}). \texttt{./run\_all.sh} then redraws every figure, table and number from the new data, and its checksum check names the outputs that changed. Sampling makes single episodes differ between reruns; the paired intervals are the quantity we expect to reproduce.
\par\endgroup

\end{document}